\documentclass[fleqn,usenatbib]{mnras}

\usepackage{subcaption}
\usepackage{amsmath}
\usepackage{amssymb}

\usepackage{newtxtext,newtxmath}
\usepackage[T1]{fontenc}
\usepackage{multirow}

\DeclareRobustCommand{\VAN}[3]{#2}
\let\VANthebibliography\thebibliography
\def\thebibliography{\DeclareRobustCommand{\VAN}[3]{##3}\VANthebibliography}

\usepackage{graphicx}

\title[bending delay in PSR-BH binaries]{A study of the light bending phenomenon under full general relativity for a pulsar in a binary with a Schwarzschild black hole}

\author[Debnath et al.]{
Jyotijwal Debnath,$^{1, 2}$\thanks{E-mail: jdebnath@imsc.res.in}
Manjari Bagchi,$^{1,2}$
and
Avishek Basu$^{3}$
\\
$^{1}$The Institute of Mathematical Sciences, C. I. T. campus, Taramani, Chennai, 600113, India\\
$^{2}$Homi Bhabha National Institute, Training School Complex, Anushakti Nagar, Mumbai 400094, India\\
$^{3}$Jodrell Bank Centre for Astrophysics, Department of Physics and Astronomy, University of Manchester, Manchester M13 9PL, UK
}

\date{Accepted XXX. Received YYY; in original form ZZZ}

\pubyear{2023}

\begin{document}
\label{firstpage}
\pagerange{\pageref{firstpage}--\pageref{lastpage}}
\maketitle

\defcitealias{Rafikov2005}{RL06}
\defcitealias{dk1995}{DK95}

\begin{abstract}
The values of the bending delays in the signal of a radio pulsar in a binary with a stellar mass black hole as a companion have been calculated accurately within a full general relativistic framework considering the Schwarzchid spacetime near the companion. The results match with the pre-existing approximate analytical expressions unless both of the orbital inclination angle and the orbital phase are close to $90^{\circ}$. For such a case, the approximate analytical expressions underestimate the value of the bending delay. {On the other hand, for systems like the double pulsar, those expressions are valid throughout the orbital phase, unless its inclination angle is very close to $90^{\circ}$}. For a pulsar$-$black hole binary, the bending phenomenon also increases the strength of the pulse profile and sometimes can lead to a small low intensity tail. 
\end{abstract}


\section{Introduction}
\label{sec:intro}
Nearly ten percent of the presently known radio pulsars are part of binary systems. Binaries for which the companion of the pulsar is also a compact star, e.g., either a white dwarf or another neutron star or a stellar mass black hole, can be used to test theories of gravity, e.g., general relativity, scalar-tensor theory, tensor-vector-scalar theory, etc \citep{Stairs2003}. The first test of general relativity via the indirect detection of gravitational waves was done by the first discovered binary pulsar PSR B1913$+$16, which is a double neutron star system  \citep{ht75}. Afterward, strong field tests of gravity were followed using the timing data of this and other double neutron star systems \citep{wh16, ksm21, cbb22}. In some cases, tests of gravity theory were possible for pulsar$-$white dwarf binaries too \citep{fwe12, afw13, ddf20, cfr20}. It is expected that pulsar$-$black hole (stellar mass) binaries would be useful to test the {black-hole `no-hair theorem'  \citep{wk99}},  `Cosmic Censorship Conjecture', to detect dipolar gravitational waves (allowed by some gravity theories) if those exist, to test the validity of non-conservative theories of gravity that predict a self acceleration of the centre of mass of the binary, etc \citep{bt14}. Although no pulsar$-$black hole binaries have been discovered yet, various theoretical studies demand that such pulsar$-$black hole binaries must exist in the Galaxy \citep{lpp94, ss03, ppr05, kph09, fcl11, cds13, css21}. The LIGO detection of gravitational waves emitted by the merger of a neutron star and a black hole in a binary supported this claim \citep{aaa20}. As the new generation of sensitive radio telescopes like MeerKAT \citep{rfg22, cfr23} and FAST \citep{han21, pan21} are discovering many pulsars presently and more discoveries are expected by them as well as by the upcoming telescopes like SKA \citep{hpb14}, it is possible that the discovery of a pulsar$-$black hole binary will happen anytime soon. Hence, it is a good time to understand how different aspects of the strong field gravity would manifest in the data of the pulsar in such a binary.

The tests of gravity are possible through `pulsar timing analysis', which is basically calculating the rotational phases of the pulsar correctly by modelling the delays between  successive `Time of Arrivals (ToAs)' of the pulses. There are different types of delays specific to the binaries, e.g, the binary R\"omer delay, the binary Shapiro delay, the binary Einstein delay, the delay due to the light bending effects, etc. bearing imprints of various aspects of gravity, including the effects of the motion of the pulsar in its orbit as well as the spacetime curvature near the companion \citep{lorimer04, kek11}. There are other delays, some of which are related to gravitational phenomena, but unrelated to the binary, like the solar system Shapiro delay, the solar system R\"omer delay, and the solar system Einstein delay. There are some other delays that are even not related to gravitational phenomena, like the dispersion delay, the aberration delay, etc. Tests of gravity usually come from modelling and measuring of gravity induced delays in binaries. As full general relativistic formalism for these gravitational delays are not easy, these are usually modelled by approximate theories like the post-Newtonian theory. However, it is not very well understood whether these approximate models for these delays will be sufficient for a pulsar$-$black hole binary. In the present paper, we study the delay due to the light bending effect under full general relativity for a pulsar in a binary with a non-rotating, stellar mass black hole as a companion. We also explore an additional manifestation of the light bending phenomenon, e.g., the change in the pulse shape.

This paper is organised as follows, in section \ref{sec:analytic}, we describe the analytical formalism of our work that includes modelling the beam of the pulsar, the geometrical description of the binary orbit, calculation of the light bending delay, and the distortion of the beam due to the light bending effect. In section \ref{sec:numerical}, we quantify our formalism with numerical results. After testing and comparing our results with earlier approximate calculations for the double pulsar, we calculate the amount of bending delays for hypothetical pulsar$-$black hole binaries. We demonstrate how the values of various parameters of such binaries affect the amount of the bending delay. Then, we show how the bending phenomenon distorts the beam and as a result, change the shape of the pulse profile. We conclude our paper with a summary and discussion in section \ref{sec:conclu}.

\section{Analytical set up}
\label{sec:analytic}

\subsection{Constructing the beam of the pulsar}
\label{subsec:beam}
The electromagnetic beam from a pulsar has many complex features. \cite{rc69} first proposed the hollow cone model of the pulsar beam, which was limited in explaining the double and single profiles only. Further, to explain triple profiles, \cite{Backer1976} added a pencil-beam around the magnetic axis. However, some of the pulsars have 4-component or 5-component pulse profiles, and to explain those, \cite{Rankin83} proposed that the cone component of the beam splits into two, the inner cone and the outer cone. However, this simple core-double cone model fails to reproduce more complex profiles, that can be generated using other models like the patchy-beam model \citep{lm88}, the fan-beam model \citep{wpz14}, etc. 

In this work, we use the simple core-double cone model. To construct these components, one needs to know the values of the half-opening angles at two locations for each of the components, usually at the location where the intensity is the maximum and a location where the intensity is some fraction of the maximum. These half-opening angles depend on the spin period ($P_s$ in s) of the pulsar as well as on the observing frequency $\nu$. As 1.4 GHz is one of the most commonly used frequency for pulsar observations, following \cite{gks93}, we use the expressions for half-opening angles for different components at $\nu=1.4$ GHz as:
\begin{subequations}
\label{eq:halfopeningangle}
\begin{align}
    \label{eq:peakcore}
    \rho_{\rm 1.4,peak}^{\rm core}= 0 ~,
\end{align}
\begin{align}
    \label{eq:peakin}
    \rho_{\rm 1.4,peak}^{\rm in}=3.7^\circ \, {\rm P_s}^{-0.5}~,
\end{align}
\begin{align}
    \label{eq:peakout}
    \rho_{\rm 1.4,peak}^{\rm out}=4.6^\circ \, {\rm P_s}^{-0.5}~ ,
\end{align}
\begin{align}
    \label{eq:halfopeninganglecore}
    \rho_{\rm 1.4, 50}^{\rm core}= 1.2^\circ \, {\rm P_s}^{-0.5}~ ,
\end{align}
\begin{align}
    \label{eq:halfopeningangleincone}
    \rho_{1.4,1 0}^{\rm in} = 4.42^\circ \, {\rm P_s}^{-0.5} ~, 
\end{align}
\begin{align}
    \label{eq:halfopeningangleoutcone}
    \rho_{1.4, 10}^{\rm out} = 5.83^\circ \, {\rm P_s}^{-0.5} ~,
\end{align}
\end{subequations} where $\rho_{\rm 1.4, peak}^{\rm core}$, $\rho_{\rm 1.4, peak}^{\rm in}$ and $\rho_{\rm 1.4, peak}^{\rm out}$ are the half opening angles corresponding to the location of the maximum intensities for the core, the inner cone, and the outer cone, respectively. Similarly, $\rho_{1.4, 10}^{\rm in}$ and $\rho_{1.4, 10}^{\rm out}$ are the half opening angles corresponding to the location where the intensity is $10\%$ of the maximum values, for the inner and the outer cones, respectively. $\rho_{1.4, 50}^{\rm core}$ is the half opening angle corresponding to the location where the intensity is $50\%$ of its maximum value for the core. The numerical factors and the exponents in the above expressions are either the ones reported by \cite{gks93} or slightly changed (within the error bars) to get a realistic model of the beam that will result in pulse profiles observed for several pulsars (will be discussed later).

For any given half-opening angle $\rho$, the corresponding radius $\mathcal{R}$ on the cross section of the beam at a height $h$ from the centre of the pulsar is given by:
\begin{equation}
\label{eq:halfopeningangleToRadius}
\mathcal{R}=h \, \tan \rho ~.
\end{equation} 

In the present work, to solve the equation of the null geodesic as will be explained later, we need to know the starting points of the light rays in the beam, i.e., the emission heights for those. We assume that all of the light rays are generated at a height $h_{\nu, {\rm em}}$ given by \cite{lorimer04}:
\begin{equation}
\label{eq:emheight}
     h_{\nu, {\rm em}} = 400 \, \text{km}\, \Big (\frac{\nu}{10^{-3}} \Big )^{-0.26} \Big(\frac{\dot{P}_{\rm s}}{10^{-15}}\Big)^{0.07} \, \Big (\frac{P_s}{\text{s}}\Big )^{0.30} ~ ,
\end{equation} where $\dot{P}_{\rm s}$ is the rate of change of $P_{\rm s}$ in the unit of ${\rm s \, s^{-1}}$. Using eqs. (\ref{eq:halfopeningangle}), (\ref{eq:halfopeningangleToRadius}) and (\ref{eq:emheight}), at the emission height at 1.4 GHz, i.e., at $h_{1.4, {\rm em}} $, one can calculate the radius corresponding to the maximum intensity of each of the components as well as the radii at $10\%$ of the maximum intensity levels for the conal components and at $50\%$ of the maximum intensity level for the core component. We study the intensity distribution on the cross-section of the beam at $h_{1.4, {\rm em}}$.

We assume that the intensity distribution for each of component of the beam is Gaussian on the cross-section of the beam, given by: 
\begin{equation}
\label{eq:gauss}
 I (r) = I_0 \rm{exp} \left( -\frac{(r-\mu)^2}{2\sigma^2} \right) ~,
\end{equation} where $\mu$ is the mean and $\sigma$ is the standard deviation of the Gaussian distribution, $r$ is the radial coordinate on the cross-section of the beam, and $I_0$ is a multiplicative factor, whose unit determines the unit of the intensity. The values of $\mu$ for a particular component is chosen equal to the value of the radius where the intensity is the maximum, obtained from eqs. (\ref{eq:halfopeningangle}) and (\ref{eq:halfopeningangleToRadius}). Hence, the value of $\mu$ for the core component is zero. The value of $\sigma$ for a particular component is chosen in such a way that the value of $r$ at which the intensity would fall $\beta\%$ of the maximum intensity would match with the value $\mathcal{R}_{\beta}$ obtained from eqs. (\ref{eq:halfopeningangle}) and (\ref{eq:halfopeningangleToRadius}). The values of $I_0$ can be chosen the same or different for the three components. This will lead to different shapes of the pulse profile. After getting the analytical intensity distributions of all three components, we generated a number of light rays on the cross-section of the beam whose surface density match with these distributions. The details of this numerical construction has been discussed in section \ref{subsec:numerical_distort}.

\begin{figure}
	\includegraphics[width=100mm]{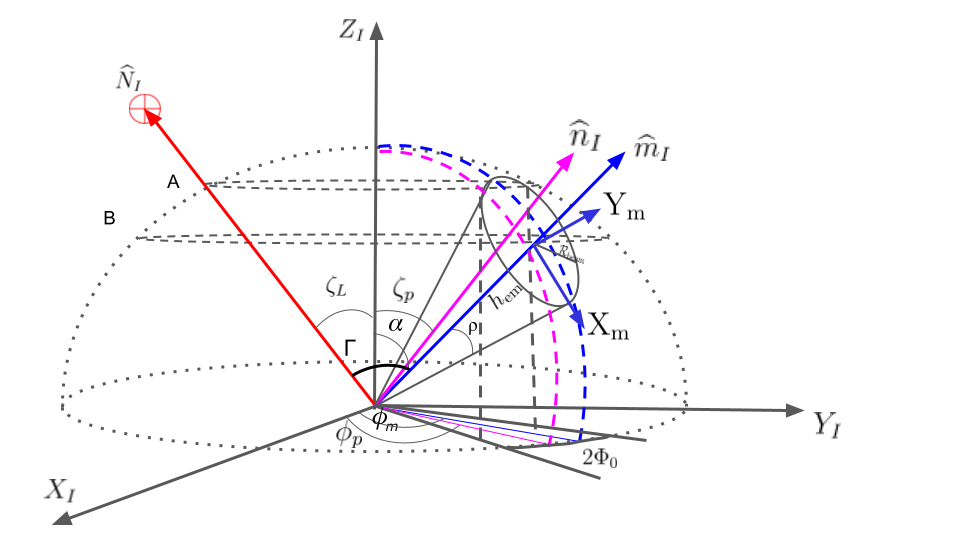}
    \caption{The beam of a pulsar and the Line-of-Sight (LoS). The meaning of various vectors and angles have been explained in the text. }
    \label{fig:spin_axis_frame}
\end{figure}

After constructing the beam, we consider the orientation of the beam with respect to various directions. Fig. \ref{fig:spin_axis_frame} shows a schematic diagram of the geometry of a pulsar and its beam at an arbitrary time $t$. For the sake of simplicity, we have shown only the radius at which the intensity of the outer cone becomes ten percent of its peak value at $\nu=1.4$ GHz, as the edge of the cone, i.e., $ \mathcal{R}_{\rm beam}= \mathcal{R}_{1.4, 10}^{\rm out}$. The corresponding half-opening angle $ \rho_{1.4, 10}^{\rm out}$ has been simply denoted by $\rho$ in this figure. ${\rm X_I} {\rm Y_I} {\rm Z_I}$ (known as the `I-frame') is a Cartesian coordinate system in which the spin axis of the pulsar is along the ${\rm Z_I}$ direction and the magnetic axis rotates around it with a period $P_{\rm s}$ making an angle $\alpha$ which is considered to remain unchanged over time. As shown in Fig. \ref{fig:spin_axis_frame}, $\widehat{m}_{\rm I}$ is a unit vector along the magnetic axis, i.e., along the axis of the beam whose half opening angle is $\rho$. The trajectory of $\widehat{m}_{\rm I}$ around the ${\rm Z_I}$-axis is shown with a dashed line marked B in the figure. All the light rays within the beam also rotate around ${\rm Z_I}$ with the period $P_{\rm s}$. The azimuthal angle of the magnetic axis, i.e., the angle between the projection of $\widehat{m}_{\rm I}$ on the ${\rm X_I} {\rm Y_I}$ plane and the ${\rm X_I}$ axis is $\phi_m$, which varies with time and changes by an amount of $2 \pi$ radian in time $P_{\rm s}$. The azimuthal angle is also called as the `longitude' or sometimes as the `phase'. While doing the numerical calculations, we have assumed that at the instant $t=0$, $\widehat{m}_{\rm I}$ is in the $\rm X_IZ_I$ plane. Hence, we can write $ \phi_m(t) = \frac{2\pi}{P_{\rm s}} t$.

The unit vector along the line-of-Sight (LoS) is denoted by $\widehat{N}_{\rm I}$. $\widehat{N}_{\rm I}$ lies on the ${\rm Z_I} {\rm X_I}$ plane and makes an angle $\zeta_{L}$ with the ${\rm Z_I}$ axis (co-latitude) as shown in the figure. We define the positive direction of the LoS from the pulsar to the observer. The angle between the LoS and $\widehat{m}_{\rm I}$ is denoted by $\Gamma$. For a particular pulsar, angles $\alpha$ and $\rho$ do not vary with time while $\Gamma$ and $\phi_m$ do. The angle $\zeta_{\rm L}$ can vary very slowly due to the free precession of the pulsar due to its ellipticity \citep{jones12}. We have ignored this effect in the present work.

Among all the light rays in the beam, only the ones having the value of the co-latitude as $\zeta_{L}$ will be aligned with the LoS once during their full rotation and will be visible (without the bending). The trajectory of these light rays is shown with a dashed line marked A in the figure. This is also the trajectory of the LoS in the rest frame of the pulsar. At any given instant t, different light rays following the trajectory A have different phases. 

However, all of the light rays in the beam will experience bending if there is a gravitating object nearby, e.g., the binary companion. In Fig. \ref{fig:spin_axis_frame}, we have shown the unit vector $\widehat{n}_{\rm I}$ along one generic light ray that has the phase $\phi_p$ and the co-latitude $\zeta_p$. The unit vectors $\widehat{m}_{\rm I}$ and $\widehat{n}_{\rm I}$ can be written as $\widehat{m}_{\rm I}=[\cos\phi_m  \sin\alpha, \, \sin\phi_m \sin\alpha, \, \cos\alpha]$ and $\widehat{n}_{\rm I}=[\cos\phi_p  \sin \zeta_p, \, \sin\phi_p  \sin\zeta_p, \, \cos  \zeta_p]$. This light ray with the co-latitude $\zeta_p$ will never fall on the LoS without bending and would not be visible unless due to the bending it seems to originate with a co-latitude $\zeta_L$. This point will be discussed in details later. 

From Fig. \ref{fig:spin_axis_frame}, it is obvious that the pulsar will be observable only if there are some light rays in the beam for which the angle $\zeta_L$ satisfy the condition $ | \alpha - \rho | \leq | \zeta_L | \leq | \alpha + \rho| $. The duration of the observability or the width of the pulse is determined by the time for which the LoS remains within the beam, which is given by the condition $\Gamma \leq \rho$. 
One additional frame is shown in Fig. \ref{fig:spin_axis_frame}. This is the ${\rm X_ m Y_ m  Z_m}$ frame or the `m-frame'. The origin of this frame is chosen at the emission height ($h_{\nu, {\rm em}}$) on the magnetic axis. The ${\rm Z_m}$-axis is taken along the magnetic axis and the ${\rm X_m Y_m}$ plane is the plane of the cross section of the beam. The intensity distributions discussed earlier can be taken on the ${\rm X_m Y_m}$ plane. The angle between the ${\rm Z_I}$-axis and the ${\rm Z_m}$-axis is $\alpha$. The direction of the ${\rm X_m}$-axis is such that if the m-frame is  shifted to the origin of the I-frame keeping the orientation of the axes unchanged, the projection of the ${\rm X_m}$-axis on the ${\rm X_ I Y_ I}$ plane would make an angle $\phi_m$ with the ${\rm X_I}$-axis.

The shape of the pulse comes from the intensity distribution over the line through which the LoS cuts the cross-section of the beam, i.e., the ${\rm X_m} \, {\rm Y_m}$ plane. If $\Phi_0$ is the half width of the pulse observed, then $2 \Phi_0$ is the difference in the phases at the point where the LoS enters the beam and the point where the LoS exits the beam (in the rest frame of the pulsar) and is shown in Fig. \ref{fig:spin_axis_frame}. In the I-frame, $2 \Phi_0$ is the difference between the azimuthal angles of the magnetic axis when the LoS enters the beam and exits the beam. 

As the m-frame can be obtained from the I-frame by a rotation by an angle $\alpha$ about the $\rm Y_I$-axis and then a rotation by an angle $\phi_m$ about the $\rm Z_I$-axis, any unit vector $\widehat{k}_{\rm m}$ in the m-frame can be converted to a unit vector $\widehat{k}_{\rm I}$ in the I-frame as:
\begin{equation}
\label{eq:unitvecINbANDm}
 \widehat{k}_{\rm I} =  R_z(\phi_m) R_y(\alpha) \widehat{k}_{\rm m} ~.
\end{equation} We use the notations $R_z(\theta)$ and $R_y(\theta)$ for active rotation matrices with the rotation angle $\theta$, about the z-axis and the y-axis, respectively, of any right handed Cartesian coordinate system. If the unit vector along the ${\rm Z_m}$-axis is denoted by $\widehat{z}_{\rm m}$ and the position vector of the starting point of a light ray on the ${\rm X_m} \, {\rm Y_m}$ plane is given by ${\vec{\mathcal{R}}_{\rm m}}$, then the unit vector along the light ray in the m-frame is given by:
\begin{equation}
\label{eq:nm}
\widehat{n}_{\rm m} = (h_{\nu, {\rm em}} \widehat{z}_{\rm m}+ {\vec{\mathcal{R}}_{\rm m}} ) (|h_{\nu, {\rm em}} \hat{z}_m+ {\vec{\mathcal{R}}_{\rm m}}|)^{-1} ~.
\end{equation} The vector ${\vec{\mathcal{R}}_{\rm m}}$ can be obtained if the coordinates of the light rays in the ${\rm X_m} \, {\rm Y_m}$ plane are known and vice versa. Using ${\vec{\mathcal{R}}_{\rm m}}$, we find $\widehat{n}_{\rm m}$ using eq. (\ref{eq:nm}) and then from $\widehat{n}_{\rm m}$, we get $\widehat{n}_{\rm I}$ using eq. (\ref{eq:unitvecINbANDm}). Note that, in Eq. (\ref{eq:nm}),  $\widehat{n}_{\rm m}$ is defined at the centre of the pulsar (where $\widehat{n}_{\rm I}$ is also defined ) not at the height $h_{\rm \nu, em}$, where the light rays are generated.

\subsection{The geometry of the pulsar in its orbit and various coordinate systems}
\label{subsec:coordinates}
\begin{figure}
	\includegraphics[width=94mm]{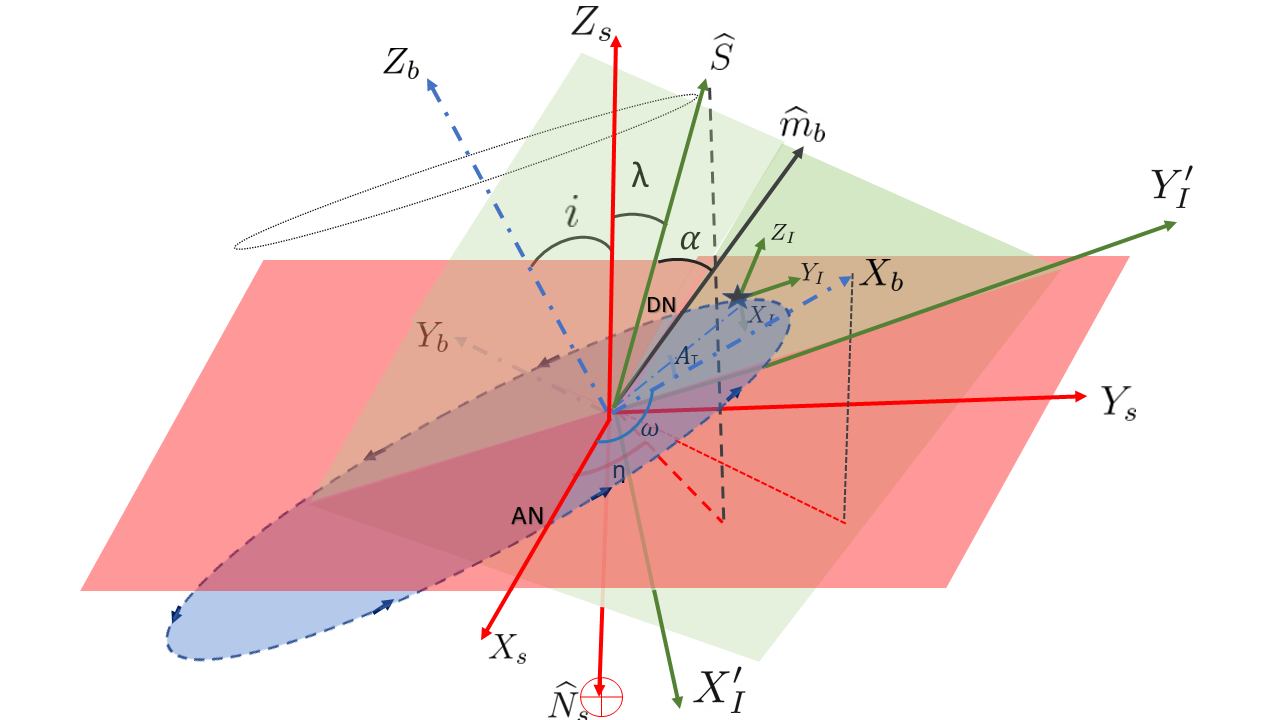}
    \caption{The orbital geometry of a pulsar (marked by a $\star$) in a binary system. Various coordinate axes and angles (as discussed in the text) are also shown. The sky plane (the ${\rm X_s}{\rm Y_s}$ plane) is shown with a light red colour and the orbital plane (the ${\rm X_b Y_b}$ plane) is shown with a light blue colour. The green plane is the ${\rm X_I^{\prime}  Y_I^{\prime}}$ plane which is parallel to the ${\rm X_I  Y_I}$ plane defined in section \ref{subsec:beam}. The unit vector along the spin axis is denoted by $\widehat{S}$, which is physically along the ${\rm Z_I}$-axis, but here shown along the ${\rm Z_I^{\prime}}$-axis, i.e., has been shifted parallely to the centre of the ${\rm X_b Y_b Z_b}$ frame. The ascending node has been denoted by AN and the descending node by DN.}
    \label{fig:geometry}
\end{figure}

We need to use some additional coordinate systems to study the pulse profile of a pulsar in a binary system. Among these, two most important coordinate systems and various angles for a binary pulsar are shown in Fig \ref{fig:geometry}. Here, ${\rm X_s Y_s Z_s}$ is a reference frame at the barycentre of the binary and the ${\rm Z_s}$-axis is along the negative direction of the LoS, i.e., from the observer to the barycentre\footnote{For a single pulsar, the LoS is defined from the pulsar to the observer and for a binary pulsar, the LoS is defined from the barycentre of the binary to the observer. However, these two directions are practically the same as in any reasonable scenario, the observer is located at a large distance compared to the size of the orbit.}. Hence, the unit vector along the LoS in the ${\rm X_s Y_s Z_s}$ frame can be represented as $\widehat{N}_{\rm s}=[0, 0, -1]$. The ${\rm X_s}$-axis is the line of the ascending node (AN) and the ${\rm X_s}{\rm Y_s}$ plane is the plane of the sky. The ${\rm X_s Y_s Z_s}$ frame is called as the `sky-frame' or the `s-frame'. ${\rm X_b} {\rm Y_b} {\rm Z_b}$ is another reference frame in which the pulsar is orbiting its companion in the ${\rm X_b Y_b }$ plane and the ${\rm X_b}$-axis is along the periastron of the orbit. Hence, the ${\rm Z_b}$-axis is along the orbital angular momentum. The ${\rm X_b} {\rm Y_b} {\rm Z_b}$ frame is also known as the `b-frame'. If $\omega$ and $i$ are the longitude of the periastron and the inclination of the orbit, respectively, then the angle between the ${\rm Z_s}$-axis and the ${\rm Z_b}$-axis is $i$ and the angle between the ${\rm X_s}$-axis and the ${\rm X_b}$-axis is $\omega$ as shown in Fig. \ref{fig:geometry}. The sky plane is shown with a light red colour and the orbital plane is shown with a light blue colour in Fig \ref{fig:geometry}.

We define two additional frames that are not shown in the figure. These are the ${\rm X_p Y_p Z_p}$ and the ${\rm X_c Y_c Z_c}$ frames, at the centres of the pulsar and its companion, respectively. We call these frames as the `p-frame' and the `c-frame'. These two frames are connected to the b-frame through parallel shifts along the line joining the pulsar and the companion. As the b-frame, the p-frame, and the c-frame are parallel to each other, any direction vector calculated in one frame will be the same in all other frames. Moreover, as already explained in footnote 1, the direction of the LoS is the same in all these three frames.

The I-frame is the same as that defined in section \ref{subsec:beam}. Here we introduce another frame, the ${\rm X_I^\prime Y_I^\prime Z_I^\prime}$ frame, which is obtained by shifting the ${\rm X_I Y_I Z_I}$ frame to the barycentre along the line joining the pulsar and the companion in such a way that the ${\rm X_I^\prime}$ axis is parallel to the ${\rm X_I}$ axis, ${\rm Y_I^\prime}$ axis is parallel to the ${\rm Y_I}$ axis, and ${\rm Z_I^\prime}$ axis is parallel to the ${\rm Z_I}$ axis. The ${\rm X_I^\prime Y_I^\prime}$ plane is shown with a green colour in Fig. \ref{fig:geometry}. The ${\rm Y_I^\prime}$-axis lies on the ${\rm X_s \, Y_s}$ plane. Note that, in case of a pulsar$-$black hole binary, the barycentre is inside the black hole as it is much heavier than the pulsar, even for a stellar mass black hole. Hence, the b-frame and the c-frame are almost the same. 

We also assume that the spin axis of the pulsar, which is along the ${\rm Z_I}$-axis, makes an angle $\lambda$ with the ${\rm Z_s}$-axis and the  projection of the spin axis in the sky plane makes an angle $\eta$ with the ${\rm X_s}$-axis. It is obvious that the angle $\lambda$ is related to $\zeta_L$ defined in the earlier section as $\lambda = 180^{\circ} - \zeta_L$. The unit vector along the spin axis is denoted by $\widehat{S}$, which is physically along the ${\rm Z_I}$-axis, but mathematically can be shifted parallelly and taken along the $Z_I^{\prime}$-axis, as done in Fig. \ref{fig:geometry}. The true anomaly $A_T$ of the pulsar is also shown in Fig. \ref{fig:geometry}. The angle $A_T+\omega$ is the orbital phase of the pulsar. The pulsar has been marked with a black star in the figure.

The total angular momentum of the system can be considered along the direction of the orbital angular momentum, i.e., along the ${\rm Z_b}$-axis. So, due to the spin-orbit coupling, the spin axis of the pulsar would precess around the ${\rm Z_b}$-axis keeping the angle between them invariant. This is known as the geodetic precession. We incorporate the geodetic precession using the well known formalism of \cite{jwt02}. Note that, this precession of the spin would make the angles $\eta$ and $\lambda$ change with time. These angles oscillate between corresponding maximum and minimum values with a period which is the same as the period of the geodetic precession. However, as we will see later, the changes in the values of these angles in one orbital period are very small.

To study the effect of the light bending, we need to find the directions of $\widehat{N}_{\rm s}$, $\widehat{n}_{\rm I}$, and $\widehat{m}_{\rm I}$ in the b-frame. If the unit vectors in the b-frame along the above three vectors are denoted by $\widehat{N}_{\rm b}$, $\widehat{n}_{\rm b}$, and $\widehat{m}_{\rm b}$, then from the geometry shown in Fig. \ref{fig:geometry}, we can write:
\begin{equation}
\label{eq:m_in_b_frame}
    \widehat{m}_{\rm b} = R_z(-\omega)R_x(-i)R_z(\eta)R_y(\lambda) \, \widehat{m}_{\rm I} ~,
\end{equation} 
\begin{equation}
\label{eq:n_in_b_frame}
    \widehat{n}_{\rm b} = R_z(-\omega)R_x(-i)R_z(\eta)R_y(\lambda) \, \widehat{n}_{\rm I}   ~,
\end{equation} and
\begin{equation}
\label{eq:LoS_in_b_frame}
    \widehat{N}_{\rm b} = R_z(-\omega)R_x(-i) \, \widehat{N}_{\rm s}  ~.
\end{equation} We use the notations $R_x(\theta)$, $R_y(\theta)$, and $R_z(\theta)$ for the active rotation matrices with the rotation angle $\theta$, about the x-axis, the y-axis, and the z-axis, respectively, of any right handed Cartesian coordinate system. In eqs. (\ref{eq:m_in_b_frame}), (\ref{eq:n_in_b_frame}), and (\ref{eq:LoS_in_b_frame}), we use the definitions of $\widehat{m}_{\rm I}$, $\widehat{n}_{\rm I}$, and $\widehat{N}_{\rm s}$ as already mentioned. Note that, as the c-frame and the b-frame are parallel to each other, all the unit vectors in these two frames are equal, i.e., $\widehat{m}_b=\widehat{m}_c$, $\widehat{n}_b=\widehat{n}_c$, and $\widehat{N}_b=\widehat{N}_c$.

\subsection{Tracing the pulsar in its orbit}
\label{subsec:orbit}
To explore the effect of the light bending phenomenon at different orbital phases of the pulsar, we track the motion of the pulsar in its orbit, i.e., in the b-frame, in a post-Keplerian manner, i.e., solve Kepler's equations to obtain the true anomaly of the pulsar in its orbit, as well as track the evolution of the orbit due to the general relativistic effects. We consider three main effects of general relativity, e.g., the rate of change of the orbital period ($\dot{P}_{\rm b}$, where $P_{\rm b}$ is the orbital period), the rate of change of the orbital eccentricity ($\dot{e}$, where $e$ is the orbital eccentricity), as well as the rate of the periastron precession ($\dot{\omega}$) \citep{lorimer04}, all up to the first order only. However, as in this work, we study the pulsar for at most one full orbit, we could in principle ignore these general relativistic effects in the orbital dynamics without affecting our results.

The position vector of the pulsar at any given time in the c-frame is denoted by $\vec{r}_{\rm c}$ and in the b-frame by $\vec{r}_{\rm b}$. $\widehat{r}_{\rm c}$ and $\widehat{r}_{\rm b}$ are unit vectors along these vectors. Hence, $\widehat{r}_c=\widehat{r}_b$ as the c-frame and the b-frame are parallel to each other.

\subsection{The light bending effect} 
\label{subsec:analytic_lightbending} 
In case of a binary pulsar, the curvature of the spacetime around the companion affects the propagation time as well as the directions of the light rays in the signal from the pulsar. The change in the direction of the light rays produces a geometric delay, as well as additional delays known as the `longitudinal bending delay'  and the `latitudinal bending delay' \citep[][hereafter RL06]{Rafikov2005} arising from the fact that the longitude and the latitude of a ray visible at any instant after bending would be different than those if there was no bending. If not modelled, these delays would manifest as orbital phase dependent delays in the timing residuals. The bending phenomenon also affects the shape of the pulse, which is a combined effect of the longitudinal and the latitudinal bendings.

It is obvious that the effect of the bending is the strongest at the superior conjunction, i.e., when the pulsar, the companion, and the earth are aligned in a straight line and the pulsar is behind the companion. This happens when the values of both of the orbital inclination angle ($i$) and the orbital phase ($\omega + A_T$) are $90^{\circ}$. However, the bending is perceptible over a range of values of $i$ and $\omega + A_T$ near $90^{\circ}$.

\subsubsection{Longitudinal bending delay} 
\label{subsec:analytic_lightbending_long}
Due to the bending, the longitudes of the light rays in the beam change. Hence, at an instant $t$, if a particular light ray had its phase (longitude) $\phi$ in the absence of bending, it would have a different phase due to the bending at the same instant $t$. On the other hand, a different light ray would have the phase $\phi$ in the presence of the bending, which had a different phase ($\phi_{\rm without}$) if the bending was not present. Thus, the difference in the phase is due to the bending is $ \Delta \phi =  \phi_{\rm without} - \phi  $. This would give rise to the longitudinal bending delay and is quantified as: 
\begin{equation}
\label{eq:exp_longdelay}
\tau_{\rm long} = \Delta \phi \times P_{\rm s} /2 \pi ~.
\end{equation} Note that, all of the light rays that falls on the trajectory of the LoS after bending, have the same value of $\Delta \phi$.

\subsubsection{Latitudinal bending delay:} 
\label{subsec:analytic_lightbending_latM} 
Due to the bending, the co-latitudes of all of the light rays also change. Hence, the light ray that originally had the co-latitude $\zeta_L$ and was visible, would no longer be so. On the other hand, another light ray that originally had a different co-latitude ($\zeta_{\rm without}$) would now have the co-latitude $\zeta_L$ and would be visible. Thus, the difference in the co-latitude due to the bending is $ \Delta \zeta =  \zeta_{\rm without} - \zeta_L$. As a result, the shape of the beam across the trajectory of the LoS would change, i.e., the half-width of the beam that is cut by the LoS ($\Phi_0$) would change. If $\Delta \Phi_0$ is the change in $\Phi_0$, then the latitudinal bending delay is quantified as:
\begin{equation}
\label{eq:exp_latdelay}
\tau_{\rm lat} = \Delta \Phi_0 \times P_{\rm s} /2 \pi ~.
\end{equation} The half width of the pulse $\Phi_0$ is already defined in sec \ref{subsec:beam} and shown in Fig. \ref{fig:spin_axis_frame}. \citetalias{Rafikov2005} gave the expression of $\Delta \Phi_0$ as:
\begin{equation}
\label{eq:delphi}
\Delta \Phi_0= \Delta \zeta \left(\frac{1}{\tan \Phi_0 \tan \zeta_L}-\frac{1}{\sin \Phi_0 \tan \alpha}\right) ~.
\end{equation} They also used the expression of $\Phi_0$ as:
\begin{equation}
\label{eq:pulse_width}
   \cos \Phi_0=\frac{\cos \rho - \cos \zeta_L \cos \alpha}{\sin \zeta_L \sin \alpha} ~ ,
\end{equation} where $\rho$ is the half-opening angle corresponding to the outer boundary of the beam. In our case, $\rho = \rho_{1.4, 10}^{\rm out}$.  Note that, eqs. (\ref{eq:delphi}) and (\ref{eq:pulse_width}) are applicable only if the distortion of the beam due to bending is negligible. As we consider the distortion of the beam, it is preferable to calculate $\Delta \Phi_0$ numerically.

For this, we first divide $P_s$ into $10^5$ time steps, and at each time, we calculate the angles between the LoS and the light rays across the boundary of the beam cross-section, i.e., the outermost light rays and note the minimum value, which corresponds to the angle between the LoS and the closest light ray at the boundary. When the LoS is far from the beam, this angle has a large value and starts to decrease as the LoS comes closer to the beam and reaches a minimum when the LoS enters the beam (as we are considering only the light rays on the edge of the beam). The value of this angle starts to increase as the LoS comes inside the beam and after sometime, when the LoS is sufficiently closer to the other edge of the beam, starts to decrease again. Eventually, it reaches a minimum again that represents the exit of the LoS from the beam. After the exit, this angle starts to increase. If $t_1$ and $t_2$ are the values of the times at which we get the two minima, i.e., the entry and exit of the LoS, then the time for which the LoS is inside the beam is given by $\Delta t= t_2 - t_1$. Using this, we calculate $2 \Phi_0=\Delta t \times \frac{2 \pi}{P_s}$.

\subsection{Earlier studies of the bending delay for a pulsar in a binary system} 
\label{subsec:analytic_bendingdelayold} 
\citet{sch90} was the first to study the bending delay for a pulsar in a binary system. However, the treatment was valid in the weak-field limit  of the edge-on binaries. Afterward, \citet[][hereafter DK95]{dk1995} studied the bending delay for a pulsar in a binary system more elaborately, and provided an approximate expression for this delay in terms of Keplerian and post-Keplerian parameters of the orbit of the pulsar as well as two additional parameters taking care of the orientation of the spin axis of the pulsar, which involves the angles $\eta$ and $\lambda$ as defined in Fig. \ref{fig:geometry}. They also explored whether the bending delay can be measurable in timing analysis of binary pulsars. However, their analysis is mostly applicable when the separation between the pulsar and the companion, projected along the sky plane is much greater than the Einstein radius of the companion. Afterward, \citetalias{Rafikov2005} overcame this limitations, and provided more accurate expressions for both of the longitudinal and latitudinal bending delays in terms of the mass of the companion, the Keplerian orbital parameters, and the angles $\eta$ and $\lambda$.

Note that, all of the above mentioned studies of the bending delays were based on the theory of gravitational lensing, in which, it is assumed that a light ray from the pulsar propagates linearly to the plane of the lens and then gets bent by an angle which is directly proportional to the mass of the gravitating body and inversely proportional to the impact parameter of the light ray. Moreover, the theory of gravitational lensing is appropriate only near the superior conjunction and should be used only when the gravitation field of the companion is not very strong. For systems like the double pulsar, the numerical results obtained from the lensing theory can lie so close to the full general relativistic results that the use of the lensing theory might not affect the results within the observationally achievable accuracy. As an example, \citet{ksm21} found that for the double pulsar, the difference between the values of the longitudinal bending delay obtained by the full general relativistic calculation and the lensing based approximation of  \citetalias{dk1995}  is about 50 nanosecond near the superior conjunction. Note that, for the double pulsar, the data near the superior conjunction is contaminated by the magnetospheric eclipse. Hence, the discrepancy of 50 nanosecond is ignorable. For a system like a pulsar$-$black hole binary, there would be no such eclipse. Moreover, the light rays pass close enough to the black hole, and hence, the gravitational field is extremely strong. In such cases, the theory of gravitational lensing might not provide accurate enough results, especially when one would expect better timing accuracy with the aid of the the enhanced sensitivity of the future instruments.

In the present work, we study bending delays without approximations mentioned in the above paragraph. We use the formalism prescribed by \citet{chandrasekhar84}, in which the motion of the gravitating body (here the companion of the pulsar) is ignored while the light rays travel through the curved spacetime. Our main aim is to study the bending delay for the pulsar$-$black hole binaries. For a canonical pulsar$-$black hole binary with the values of various parameters as displayed in Table \ref{tab:PSRBH}, the semi-major axis of the relative orbit is 0.15 AU. So, even if the light-ray travels along the largest possible path across the binary, it will take only about 1.2 minutes for the light ray to travel the distance between the pulsar and the companion, while the orbital period of the binary is 7920 minutes. Hence, in 1.2 minutes, the displacement of the companion is negligible, as for a pulsar$-$black hole system, the companion is much heavier than the pulsar, i.e., the orbit of the companion is significantly smaller. Hence, the use of the formalism of \citet{chandrasekhar84} would not give any perceptible error. In addition to the bending delays, we also study how the light bending phenomenon distort the beam and as a result change the shape of the profile.

\subsection{Full general relativistic treatment of the bending} 
\label{subsec:analytic_bendingdelay} 
To study the effect of bending, we need to define a couple of additional reference frames. As these frames are defined with the help of $\widehat{n}_{\rm b}$, i.e., the unit vector along a light ray (before bending) in the b-frame, these frames are unique to each of the light rays. In this section, we discuss the formalism of the light bending for one generic light ray, however, it will be applicable for each of the light rays in the beam.

\begin{figure}
	\includegraphics[width=100mm]{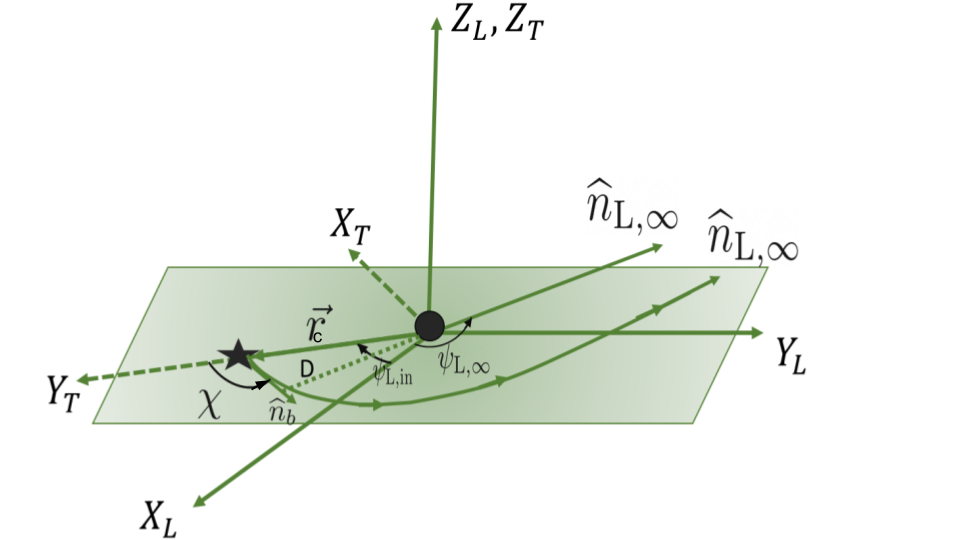}
    \caption{A schematic diagram showing the bending of a light ray when the light ray at the time of emission travels towards a gravitating object. Here, the source of the light ray is the pulsar, shown by a black star and the gravitating object is the binary companion of the pulsar shown by a black filled circle. Various axes and angles are explained in the text.}
    \label{fig:f3}
\end{figure}

We know that in the Schwarzchild metric, a light ray always moves in a plane known as the plane of the light ray. This plane must contain the pulsar as it is the starting point of the light ray. This plane also contains the companion as it is the gravitating object about which the bending of the light takes place. These facts imply that the plane of the light ray contains $\vec{r}_{\rm c}$ and $\vec{n}_{\rm c}$, or, in other words, $\vec{r}_{\rm b}$ and $\vec{n}_{\rm b}$. Using this fact, the first frame that we define here is the ${\rm X_L} {\rm Y_L} {\rm Z_L}$ frame which is centred at the gravitating body, i.e., the companion of the pulsar and the ${\rm X_L} {\rm Y_L} $ plane is the plane of the light ray described above.  Hence, the direction of the $\rm Z_L$-axis is given by $\widehat{r}_b \times \widehat{n}_b$. We call the ${\rm X_L} {\rm Y_L} {\rm Z_L}$ frame as the L-frame and it is shown with a light green colour in Fig. \ref{fig:f3} where the gravitating body, i.e., the companion of the pulsar is shown as a black filled circle and the pulsar is shown with a filled star. The directions of the ${\rm X_L}$ and ${\rm Y_L}$ axes will be discussed later.

The light rays follow null geodesics and the equation for a null geodesic in Schwarzchild spacetime around a gravitating object in the L-frame is given by \citep{chandrasekhar84}:
\begin{equation}
    \label{eq:null_geodesics}
    \left(\frac{du}{d\psi}\right)^2=\frac{2G M_c u^3}{c^2}-u^2+1/D^2 ~, 
\end{equation} where $u$ is the inverse of the radial position of the light ray in the L-frame, $D$ is the impact parameter, which is the perpendicular distance between the initial direction of the light ray and the gravitating  body, and $\psi$ is the azimuthal position of the light ray in the L-frame, i.e., measured from the ${\rm X_L}$ axis, $M_c$ is the mass of the gravitating body (companion of the pulsar), $G$ is the gravitational constant, and $c$ is the speed of light in vacuum.

Following \citet{Poutanen_2020}, the impact parameter $D$ can be written as:
\begin{equation}
\label{eq:impact parameter}
    D=\frac{r_{\rm L} \sin(\chi)}{\sqrt{1-\frac{2 G M_c}{c^2 {r}_{\rm L}}}} ~,
\end{equation} where the position vector of the pulsar in the L-frame is denoted by $\vec{r}_{\rm L}$ and $r_L=|\vec{r}_{\rm L}|$. As the centres of the c-frame and the L-frame are the same, $\vec{r}_{\rm L}= \vec{r}_{\rm c}$.

The emission angle $\chi$ is the angle between $\vec{r}_{\rm L}$ and $\widehat{n}_b$. Depending on the value of $\chi$, two cases can arise, i.e., when $\pi/2< \chi\leq \pi$, the light ray would travel towards the companion and when $0\leq \chi\leq\pi/2$ light ray would travel away from the companion, as understandable from Fig. \ref{fig:f3}.

The gravitational bending is important only for the first case, i.e., when a light ray initially moves towards the companion. \citet{chandrasekhar84} showed that in such a case, one can get physically acceptable geodesics along which the light ray can travel from a large distance to a point near the gravitating body and then go to an infinite distance only if $D> 3\sqrt{3} G M_c / c^2$. For this condition, he also provided the parametric forms of $u$ and $\psi$ that satisfy eq. (\ref{eq:null_geodesics}) in terms of the roots of the right hand side of eq. (\ref{eq:null_geodesics}). We use these parametric forms, as this is the case when the light rays from the pulsar in a binary system can reach the observer after experiencing bending near the companion.

It is obvious that the position vector of the pulsar in the L-frame, i.e., $\vec{r}_{\rm L}$, is also the initial (at the time of emission) radius vector of the light ray in the L-frame, i.e., $u_{\rm L, in} = 1/|\vec{r}_{\rm L}|$. Using $u=u_{\rm L, in}$ in the parametric forms of $u$ and $\psi$, we get the initial value of the azimuthal angle $\psi_{\rm L, in}$. Similarly, using $u=0$ in those parametric forms, we get the final value of the azimuthal angle $\psi_{\rm L, {\infty}}$. 

For the second case, i.e, when the light rays initially move away from the companion, the light rays can escape the gravitating body for any value of the impact parameter. Obviously, for this to happen, if the gravitating body is a black hole, then the emission point of the light rays must be outside the event horizon of the black hole, which would be the case for a pulsar$-$black hole binary. Although, the bending is very weak if the light rays initially move away from the companion, we still consider these cases. However, in such a case, it is very unlikely that the light rays that reach the observer have $D\leq 3\sqrt{3}G M_c / c^2$. This fact can be understood from Fig. \ref{fig:scematic_bending}. 

Hence, in the present work, we consider only the cases when $D> 3\sqrt{3}G M_c / c^2$ regardless of the value of $\chi$. However, we only discuss the case of $\pi/2< \chi\leq \pi$ in detail and show in Fig. \ref{fig:f3}, as the bending is stronger in this situation. Note that, the older works of \citetalias{dk1995} and \citetalias{Rafikov2005} are valid only for $D>> 3\sqrt{3}G M_c / c^2$ and the orbital phase near $90^\circ$.

The final direction of the light ray in the L-frame can be written as $\widehat{n}_{\rm L,\infty}=[\cos \psi_{\rm L, \infty},\ \sin \psi_{\rm L, \infty},\ 0]$. When $\pi/2< \chi\leq \pi$, we choose $\psi_{\rm L,in}$ to be negative and when $0\leq \chi\leq\pi/2$, we choose $\psi_{\rm L,in}$ to be positive and $\psi_{\rm L,\infty}$ must be positive always. Now, we need to transform $\widehat{n}_{\rm L,\infty}$ from the L-frame to the b-frame to estimate the effect of the bending. This transformation seems difficult as we do not know the directions of the $\rm X_L$-axis and the $\rm Y_L$-axis with respect to the b-frame or other related frames like the c-frame, the p-frame, the s-frame, or the I-frame. However this transformation can be done with the help of a new frame called the ${\rm X_T Y_T Z_T}$ frame or the T-frame. We choose the ${\rm Y_T}$ axis along $\vec{r}_{\rm c}$ and the ${\rm Z_T}$-axis along the ${\rm Z_L}$-axis. 

As we know the directions of the ${\rm Y_T}$-axis and the ${\rm Z_T}$-axis, we can find the direction of the ${\rm X_T}$-axis, which is along $\widehat{y}_T \times \widehat{z}_T$, where $\widehat{y}_T$ is the unit vector along the ${\rm Y_T}$-axis and $\widehat{z}_T$ is the unit vector along the ${\rm Z_T}$-axis. It is obvious that the ${\rm X_L Y_L}$ plane and the ${\rm X_T Y_T}$ plane are the same plane, i.e., the plane of the light ray which contains $\widehat{n}_b$ and $\widehat{r}_{\rm b}$ as shown in Fig. \ref{fig:f3}. Fig. \ref{fig:f3} also shows that the ${\rm X_T Y_T Z_T}$ frame can be obtained from the ${\rm X_L Y_L Z_L}$ frame by a rotation around the ${\rm Z_L}$-axis by an amount of $-(\pi/2-\psi_{\rm L,in})$ (note that, $\psi_{\rm L,in}$ is negative in Fig. \ref{fig:f3}). Hence, $\widehat{n}_{\rm T,\infty}=R_z(\pi/2-\psi_{\rm L,in})\widehat{n}_{\rm L,\infty}$.

Next, we find a relation between the T-frame and the b-frame to transform $\widehat{n}_{\rm T,\infty}$ from the T-frame to the b-frame, i.e., to obtain $\widehat{n}_{\rm b,\infty}$. If the  angle between the ${\rm Z_T}$-axis and the ${\rm Z_b}$-axis is $\theta_{\rm Tb}$ and the angle between the ${\rm Y_T}$-axis and the ${\rm Y_b}$-axis is $\phi_{\rm Tb}$, then we can write 
\begin{equation}
\theta_{\rm Tb}=\cos^{-1}(\widehat{z}_{\rm T} \cdot \widehat{z}_{\rm b}) = \cos^{-1}(\widehat{z}_{\rm T} \cdot [0,0,1]) ~,
\label{eq:thetaTb}
\end{equation} and
\begin{equation}
\phi_{\rm Tb} = \cos^{-1}(\widehat{y}_{\rm T} \cdot \widehat{y}_{\rm b}) = \cos^{-1}(\widehat{y}_{\rm T} \cdot [0,1,0] ) ~,
\label{eq:phiTb}
\end{equation} where the unit vectors $\widehat{z}_b$ and $\widehat{y}_b$ are represented by $[0,0,1]$ and $[0,1,0]$, respectively. Using eqs. (\ref{eq:thetaTb}) and (\ref{eq:phiTb}), we can get the values of $\theta_{\rm Tb}$ and $\phi_{\rm Tb}$ as $\widehat{z}_T$ and $\widehat{y}_T$ are known.

Using these angles, we get the final direction of the light ray in the b-frame as:
\begin{equation} 
\widehat{n}_{\rm b,\infty}= R_z(\phi_{\rm Tb}) R_y(\theta_{\rm Tb}) \, \widehat{n}_{\rm T, \infty} ~.
\label{eq:nbT}
\end{equation}

To estimate the values of the bending delays, first, we note a particular time, at which, in the absence of the bending, a light ray is aligned with $\widehat{N}_{\rm I}$, i.e., with the LoS. As obvious from Fig. \ref{fig:spin_axis_frame}, at this time, the phase of the light ray is zero and the co-latitude is $\zeta_L$ as it is aligned with $\widehat{N}_{\rm I}$. Now, we identify the light ray which would have the phase zero and the co-latitude $\zeta_L$ at that particular time in the presence of bending. We do this with the help of the final directions of the light rays after bending, i.e., $\widehat{n}_{\rm b,\infty}$ as described above. 

After this, we note what was the phase and the co-latitude of this particular light ray at that particular time when the bending was absent. If these values were $\phi_{\rm old}$ and $\zeta_{\rm old}$, then we can write $\Delta \phi = \phi_{\rm old}$ and $\Delta \zeta = \zeta_{\rm old }- \zeta_{\rm L}$. We use this value of $\Delta \phi$ in the expression of the longitudinal bending delay (eq. \ref{eq:exp_longdelay}). The value of $\Delta \zeta$ can be used in eq. (\ref{eq:delphi}) to get the value of $\Delta \Phi_0$ (in addition to eq. (\ref{eq:pulse_width})). This $\Delta \Phi_0$ can be used in eq. (\ref{eq:exp_latdelay}) to estimate the value the latitudinal delay. However, as we mentioned earlier in the last paragraph of section \ref{subsec:analytic_lightbending_latM}, we calculate $\Delta \Phi_0$ numerically whenever possible. The values of $\Delta \Phi_0$ found numerically match with the values obtained using eqs. (\ref{eq:delphi}) and (\ref{eq:pulse_width}) unless the bending is very strong, i.e., near the superior conjunction.

Note that, as the bending is not extremely strong in our case, the light ray that gets aligned with the LoS after bending is not very far from the light ray which gets aligned with the LoS without bending. Hence, to calculate the bending delays, we check only the light rays within a small area around the original ray. 

\subsection{Distortion of the beam and the pulse shape} 
\label{subsec:analytic_distort} 
As we mentioned earlier, we generate a number of starting points of the light rays on the ${\rm X_m Y_m}$ plane following the core-double cone intensity distributions. From the coordinates of these starting points, we know their position vectors in the ${\rm X_m Y_m}$ plane and hence, using eqs. (\ref{eq:nm}), (\ref{eq:unitvecINbANDm}), and (\ref{eq:n_in_b_frame}), we get the initial direction $\widehat{n}_{\rm b}$ of each of the light rays. Then, using the formalism described in sec. \ref{subsec:analytic_bendingdelay} , we get the final directions $\widehat{n}_{\rm b,\infty}$ for each of them. Then we convert each of these $\widehat{n}_{\rm b,\infty}$ to $\widehat{n}_{\rm m,\infty}$ using the inverse transformations of eqs. (\ref{eq:n_in_b_frame}) and (\ref{eq:unitvecINbANDm}). From $\widehat{n}_{\rm m,\infty}$, we get the position vectors and the coordinates on the ${\rm X_m Y_m}$ plane of the apparent starting points of these bended light rays using eq. (\ref{eq:nm}). The surface density of these apparent starting points give the intensity distribution on the cross-section of the beam after bending.

Note that, the initial direction of the light rays change continuously as the direction of the ${\rm X_m Y_m}$ plane changes continuously with respect to the I-frame as the beam rotates about the ${\rm Z_I}$-axis. As a result, the intensity distribution on the cross section changes rapidly within a spin period if the bending effect is strong enough (as near the superior conjunction).

\section{Numerical Study}
\label{sec:numerical}

\subsection{Features of the bending delay: demonstration for the double pulsar}
\label{subsec:numerical_bendingdelay_dns}
\begin{figure*}
\begin{subfigure}[b]{\columnwidth}
	\includegraphics[width=\textwidth]{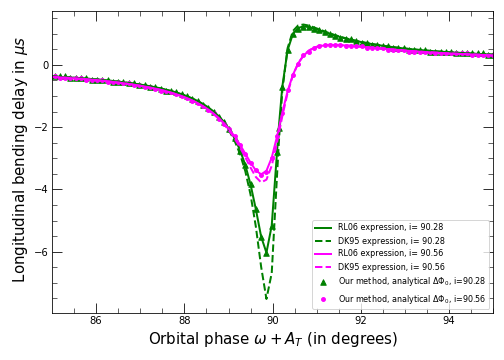}
\end{subfigure}
\begin{subfigure}[b]{\columnwidth}
\includegraphics[width=\textwidth]{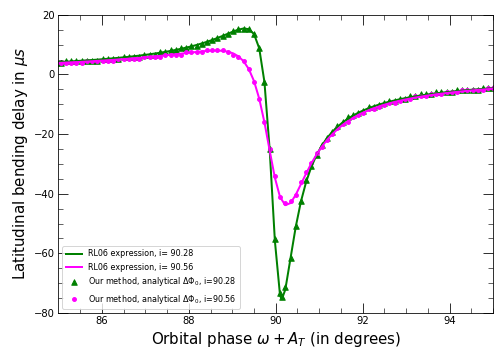}
\end{subfigure}	
    \caption{Bending delays in the signal of PSR J0737$-$3039A calculated in different methods. The left panel shows the longitudinal bending delay and the right panel shows the latitudinal bending delay. To calculate the values of the latitudinal bending delay in formalism, we have calculated $\Delta \Phi_0$ using eqs. (\ref{eq:pulse_width}) and (\ref{eq:delphi}).}
    \label{J0737_bending_delay}
\end{figure*}

To test our formalism, we first calculate the bending delays for the double pulsar, PSR J0737-3039A/B. Here we calculate the bending delay of the signal of the fast pulsar (pulsar A) due to the slow pulsar (pulsar B). We compare our results with the ones obtained by approximate equations of \citetalias{dk1995} and \citetalias{Rafikov2005}. In Fig. \ref{J0737_bending_delay}, we plot the bending delays in microseconds along the abscissa and the orbital phase $\omega+A_T$ in degrees along the ordinate. In this figure, the dashed lines are for the expression given by \citetalias{dk1995}, the solid lines are for the expressions given by \citetalias{Rafikov2005} and the circles and triangles are the results obtained by our method. Note that, \citetalias{dk1995} provided the expression only for the longitudinal bending delay. Instead of the full orbit, we plot the bending delay only around the orbital phase of $90^\circ$. In this plot, instead of updated parameters, we have used the values of the parameters for the double pulsar the same as those used by \citetalias{Rafikov2005} for the sake of comparison. Like them, we have also used two different values of the inclination angle, $i=90.56^\circ$ (magenta lines and circles) and $i=90.28^\circ$ (green lines and triangles). We see exact agreements between the results obtained by our method and the ones obtained from the expressions of \citetalias{Rafikov2005}. However, a difference between the results obtained by our method and the ones obtained from the expression of \citetalias{dk1995} is visible for $i=90.28^\circ$, and the difference is very small when $i=90.56^\circ$. We have also checked (but not plotted) that if we use the updated parameters for the double pulsar as reported by \cite{ksm21}, the results of all three expressions agree. The main reason is the fact that  \cite{ksm21} obtained $89.35^{\circ}$ (or $90.65^{\circ}$), which is further from $90^{\circ}$. These facts validate the correctness of our method, but the question remains whether the expressions of \citetalias{Rafikov2005} are applicable always.

One interesting aspect of bending is revealed in our formalism. It is the fact that the longitudinal bending delay curve would have an irregular region when both of the inclination angle and the orbital phase have their values close to $90^{\circ}$. In this irregular region, the bending delay has one or more sudden huge jump or jumps in its value with the change in the value of the orbital phase.

\begin{figure}
\centering
	\includegraphics[width=\columnwidth]{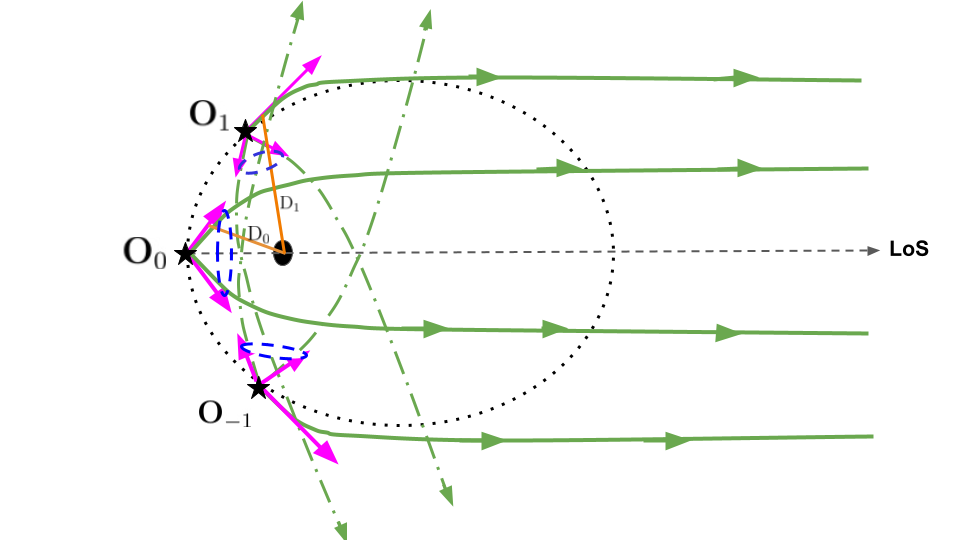}
    \caption{A schematic diagram demonstrating the light bending phenomenon in the signal of a pulsar in a binary. The companion is shown with a black filled circle and three possible position of the pulsar with black stars ($\star$). These positions are denoted by ${\rm O_{-1}}$, ${\rm O_0}$, and ${\rm O_1}$. ${\rm O_0}$ is the position of the pulsar at the superior conjunction. Initial directions of a few of the light rays are shown with magenta arrows and the green arrows are the directions of the corresponding light rays after bending. Among these, only the rays indicated by solid green lines reach the observer, the rays indicated by dashed green lines do not reach the observer. It is clear that when the pulsar is at ${\rm O_0}$, all the light rays from the circular region shown in blue (edge of a three dimensional hollow cone) reach the observer although only two such rays are shown. All these light rays experience different amount of bending as their initial directions are different. On the other hand, at positions ${\rm O_1}$ and ${\rm O_{-1}}$, which are away from ${\rm O_0}$, there is only one direction from which the observer can receive the signal from the pulsar. $D_0$ is the impact parameter of one light ray from ${\rm O_0}$ that was emitted towards the companion and after bending reaches the observer while $D_1$ is the impact parameter of one light ray from ${\rm O_1}$ that was emitted away from the companion but still reaches the observer. It is clear that the value of $D_1$ is significantly larger than that of $D_0$. That is why we consider only the cases where the value of the impact parameter is greater than $3 \sqrt{3} G M_c/c^2$, even when the light ray was initially moving away from the companion, as mentioned in sec. \ref{subsec:analytic_bendingdelay},}
    \label{fig:scematic_bending}
\end{figure}

The physical reason for this irregularity is the fact that in such a case, the rays emitted along different directions from a circular region can reach the observer as shown by the blue circle for the position ${\rm O}_0$ of the pulsar in the schematic diagram in Fig. \ref{fig:scematic_bending}. This phenomenon makes the longitudinal bending delay curve irregular. Note that, in Fig. \ref{fig:scematic_bending}, three possible orbital positions of the pulsar, ${\rm O}_{-1}$, ${\rm O}_0$, and ${\rm O}_1$ are marked. For each of these orbital positions, some of the rays emitted by the pulsar at a particular instant, i.e., at a particular rotational phase are shown with arrows. Here, we need to remember that at any given rotational phase, light rays from the full beam of the pulsar does not reach the observer, only one of the rays that is aligned with the LoS reach the observer. In Fig. \ref{fig:scematic_bending}, initial directions of a few of the light rays are shown with magenta arrows and the green arrows are the directions of the corresponding light rays after bending. Among these, only the rays indicated by solid green lines reach the observer, the rays indicated by dashed green lines do not reach the observer. The latitudinal bending delay curve would also show an irregular region when both of the inclination angle and the orbital phase have their values close to $90^{\circ}$. However, the reason for that is the distortion of the shape of the beam as will be demonstrated in section \ref{subsec:numerical_distort}.

\begin{figure*}
\begin{subfigure}[b]{0.49\textwidth}
\includegraphics[width=\textwidth]{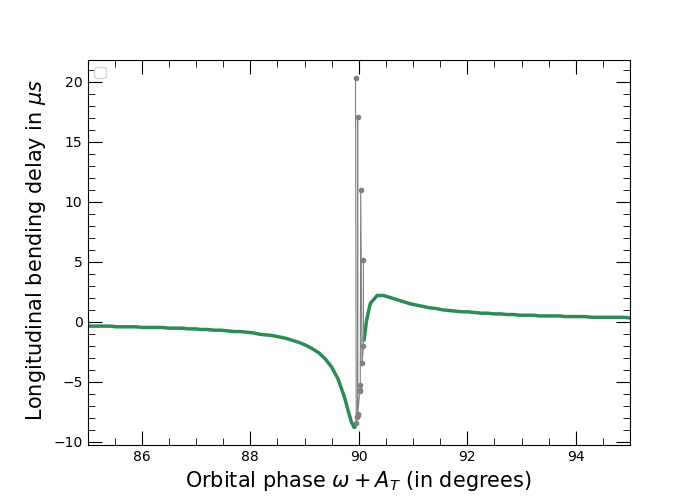}
\label{fig:J0737_long_delay_i_90_14}
\end{subfigure}
\begin{subfigure}[b]{0.49\textwidth}
\includegraphics[width=\textwidth]{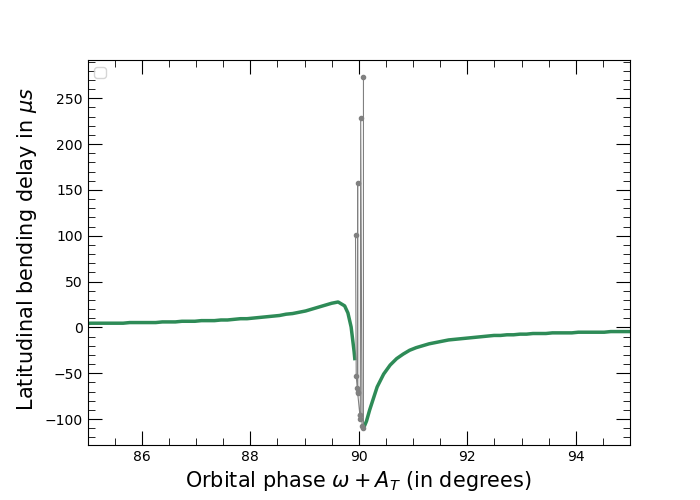}
\label{fig:J0737_lati_delay_i_90_14}
\end{subfigure}
\caption{Bending delays obtained in our method for the signal of PSR J0737$-$3039A if it had $i=90.14^\circ$. The left panned shows the longitudinal bending delay and the right panned shows the latitudinal bending delay. Both of the delay curves become irregular between $ \omega + A_T \simeq 89.8^{\circ}$ and $\omega + A_T  \simeq 90.2^{\circ}$. The irregular regions are shown in gray. Here the values of the bending delay have been calculated using our formalism and the value of $\Delta \Phi_0$ have been calculated using eqs. (\ref{eq:pulse_width}) and (\ref{eq:delphi}).}
\label{fig:J0737_bending_delay_i_90.14}
\end{figure*}

The exact value of $\omega + A_T$ at which the delay curves become irregular depends on the value of $i$. As an example, in Fig. \ref{fig:J0737_bending_delay_i_90.14}, we plot the bending delay for PSR J0737$-$3039A by setting $i=90.14^\circ$ and see that the bending delay curves become irregular between $\omega + A_T  \simeq 89.9^{\circ}$ and  $A_T + \omega \simeq 90.1^{\circ}$. Such features would be absent in the approximate formalism given by \citetalias{Rafikov2005} or \citetalias{dk1995}. 

If either of the inclination angle or the orbital phase (or both) is sufficiently different than $90^{\circ}$, the observer will see the light ray from only one particular direction of emission that depends on the values of these parameters. Hence, there would not be any irregular region in the bending delay curves as seen in Fig. \ref{J0737_bending_delay}. ${\rm O}_{-1}$ and ${\rm O}_{1}$ are two such positions in Fig. \ref{fig:scematic_bending}. Note that, at these two positions, the light rays that reach the observer were initially moving away from the companion, and hence those experience weak bending.

We need to remember the fact that, for the double pulsar, if we use the values of $\alpha=4^{\circ}$ and $\zeta_L=50^{\circ}$ as given by \citetalias{Rafikov2005} and calculate the value of $ \rho_{1.4, 10}^{\rm out}=38.69^{\circ}$ using eq. (\ref{eq:halfopeningangleoutcone}), the condition for the pulsar to be visible, $ | \alpha -  \rho_{1.4, 10}^{\rm out} | \leq | \zeta_L | \leq | \alpha +  \rho_{1.4, 10}^{\rm out} | $ is not satisfied. However, using the value of $\Phi_0=115^{\circ}$ as given by \citetalias{Rafikov2005}, as well as $\zeta_L$ and $\alpha$ in eq. (\ref{eq:pulse_width}), we get $\rho=51.73^{\circ}$, which satisfy $ | \alpha -  \rho | \leq | \zeta_L | \leq | \alpha +  \rho | $ if we assume $\rho_{1.4, 10}^{\rm out}=\rho$.  This means that either the above values of either or both of the angles $\alpha$ and $\zeta_L$ are not realistic or the beam of the pulsar A is different than that represented by eqs. (\ref{eq:halfopeningangle}). Hence, for the double pulsar, we do not create the beam and do not calculate the value of $\Delta \Phi_0$ numerically.  Instead, we calculate the values of $\Delta \phi$ and $\Delta \zeta$ by creating light rays in small regions around the LoS. We use this value of $\Delta \phi$ to calculate the longitudinal bending delay using eq. (\ref{eq:exp_longdelay}). We use the value of $\Delta \zeta$ we obtain and the values of $\Phi_0$, $\zeta_L$, and $\alpha$ as given by \citetalias{Rafikov2005} in eq. (\ref{eq:delphi}) to get the value of $\Delta \Phi_0$. We use this value of  $\Delta \Phi_0$ in eq. (\ref{eq:exp_latdelay}) to calculate the values of the latitudinal bending delay. However, eqs. (\ref{eq:delphi}) and (\ref{eq:pulse_width}) are not valid when the bending is strong and the beam is distorted. Hence, the shape of the irregular region seen in the latitudinal bending delay curve (the right panel of Fig. \ref{fig:J0737_bending_delay_i_90.14}) is not correct.  If $\Delta \Phi_0$ was calculated numerically, the shape of the irregular region would be different - we would rather see a sudden but continuous jump in the latitudinal bending delay. However, when bending is comparatively weaker, the values of the latitudinal bending delay calculated using the numerically obtained value of $\Delta \Phi_0$ match exactly with the one obtained by using the value of $\Delta \Phi_0$ calculated using eqs. (\ref{eq:pulse_width}) and (\ref{eq:delphi}). These facts will be demonstrated in the next section for a hypothetical pulsar$-$black hole binary.

\begin{figure}
\includegraphics[width=\columnwidth]{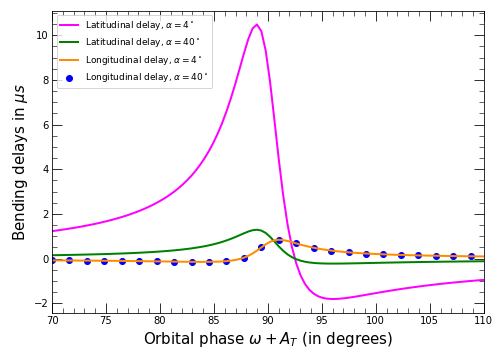}
\caption{The longitudinal and the latitudinal bending delays in the signal of a system like PSR J0737$-$3039A for two different values of $\alpha$, for $\zeta_L=50^{\circ}$, $i=87.5^\circ$, and $\eta=45^{\circ}$. Here the values of the bending delay have been calculated using our formalism and the value of $\Delta \Phi_0$ have been calculated using eqs. (\ref{eq:pulse_width}) and (\ref{eq:delphi}).}
\label{fig:bendingdelay_LOSexplore}
\end{figure}

From Fig. \ref{fig:spin_axis_frame}, it is clear that when $\zeta_L = \alpha$, the trajectories A and B coincide and the LoS cuts the beam through the midpoint. On the other hand, when $ \alpha \ne \zeta_L$, the trajectories A and B are different and the LoS cuts the beam away from its midpoint. This is the case in our calculations so far. Now we explore what happens if the LoS goes close to the midpoint of the beam, which happens when the value of $\alpha$ is close to that of $\zeta_L$. For this purpose, we change the value of $\alpha$ to $40^\circ$, and keep the values of $\zeta_L=50^{\circ}$ and $\Phi_0=115^{\circ}$ unchanged. In Fig. \ref{fig:bendingdelay_LOSexplore}, we notice the fact that the values of the latitudinal bending delay  are much smaller when the LoS cuts the beam near the midpoint ($\alpha \simeq \zeta_L$) in comparison to those when the LoS cuts the beam away from it ($\alpha \ne \zeta_L$). However, the value of the longitudinal bending delay does not depend on the location of the trajectory of the LoS across the beam. For $\alpha \ne \zeta_L$, the values of the latitudinal bending delay are much larger than those of the longitudinal bending delay while for the $\alpha \simeq \zeta_L$, these two quantities are of the same order. In the next section, we will see that for $\alpha = \zeta_L$, the value of the latitudinal bending delay is very small, even smaller than that of the longitudinal bending delay. In that section, we will even see that a pulsar$-$black hole binary with $\alpha = \zeta_L$, has smaller values of the latitudinal bending delays than those for the double pulsar with $\alpha \ne \zeta_L$. All these facts emphasise the impact of the location of the trajectory of the LoS across the beam on the value of the latitudinal bending delay, which is higher when the trajectory is far from the midpoint. This happens due to the enhanced distortion of the beam which will be clear in sec. \ref{subsec:numerical_distort}. Note that, in this section, we kept the value of $\eta$ fixed at $45^{\circ}$ following \citetalias{Rafikov2005}. The dependence of the values of the bending delays on $\eta$ will be studied in the next section for hypothetical pulsar$-$black hole binaries.

In addition to the magnitude of the bending delays, we also notice that the shapes of the bending delay curves in Fig. \ref{fig:bendingdelay_LOSexplore} are different than those in Fig. \ref{J0737_bending_delay}. The reason for this is the fact that the values of $i$ chosen  in Fig. \ref{J0737_bending_delay} are greater than 90$^{\circ}$ following \citetalias{Rafikov2005}, while in Fig.\ref{fig:bendingdelay_LOSexplore}, we have $i < 90^{\circ}$. We have checked that if all other parameters are the same, then the bending delay curves for $ i= 90^{\circ} + \Delta$ (where $\Delta$ is an arbitrary value of an angle) can be thought as a double reflection of the delay curve for $i= 90^{\circ} - \Delta$, first a vertical reflection followed by a horizontal reflection. This is the case for both of the longitudinal and the latitudinal bending delays. As in a pulsar timing analysis, one can not distinguish between two such values (equally spaced about $90^{\circ}$) of $i$, and it is a common practise to use only acute angle values for $i$, for the rest of the paper, we restrict ourselves only with the values of $i \le 90^{\circ}$.

The prescription we used, i.e., ignoring the motion of the companion of the pulsar is not physically accurate for a double neutron star system where both objects are of comparable masses. In such a case, one should employ a more sophisticated theory like the Lorentz covariant theory of light propagation in the gravitational fields of moving bodies as proposed by \citet{ks99}. However, it is not our aim to study the bending delay for double neutron star binaries, we aim to study the bending delay in the case of pulsar$-$black hole binaries where this formalism is valid. In this section, we have calculated the values of the bending delays for the double pulsar using our formalism for the sake of comparison with the results of \citetalias{Rafikov2005} and \citetalias{dk1995}.

\subsection{Bending delay in pulsar$-$black hole binaries}
\label{subsec:numerical_bendingdelay_psrbh}
Next, we compute bending delays for hypothetical pulsar$-$black hole binaries. As there is no  pulsar$-$black hole binary known at present, and different theoretical models predict different values of parameters for such binaries, we use some standard values for the parameters that are not too different than the theoretical prediction of one of the models by \citet{css21}, namely ZM-001. These parameters are the mass of the pulsar $M_p$, the mass of the companion $M_c$, $P_{\rm s}$, $\dot{P}_{\rm s}$, $P_{\rm b}$, and $e$. The values of these parameters are given in Table \ref{tab:PSRBH}. There are some other parameters whose values are needed for our calculations, those are $i$, $\omega$, $\eta$, $\lambda$, and $\alpha$. The values of $\alpha$, $\eta$ and $\lambda$ are not observationally measurable. As mentioned in section \ref{subsec:coordinates}, $\eta$ and $\lambda$ vary very slowly with time due to the geodetic precession. For the pulsar$-$black hole binary with parameters as given in Table \ref{tab:PSRBH} with $i=89^{\circ}$, over one orbit, the value of $\lambda$ changes by $-4.17\times 10^{-4}$ degrees and the value of $\eta$ changes by $-3.40\times 10^{-4}$ degree when their initial values are arbitrarily chosen as $130^\circ$ and $45^\circ$, respectively. Unless otherwise mentioned, we choose the value of $\alpha$ as $50^\circ$, and the initial values of $\lambda$ and $\eta$ as $130^\circ$ and $45^\circ$, respectively. We denote these initial values simply as $\lambda$ and $\eta$. As the effect of the bending depends strongly on the value of $i$, we use various values of $i$ for various studies as will be discussed due course.

We have mentioned in section \ref{subsec:beam} that the pulsar is observable only if there are some light rays in the beam for which the angle $\zeta_L$ satisfy the condition $ | \alpha -  \rho_{1.4, 10}^{\rm out} | \leq | \zeta_L | \leq | \alpha +  \rho_{1.4, 10}^{\rm out} | $. We also know that $\lambda = 180^{\circ} - \zeta_L$ (from Fig \ref{fig:geometry}). Using eq. (\ref{eq:halfopeningangleoutcone}), we get $\rho_{1.4, 10}^{\rm out}=4.51^{\circ}$ for the value of $P_s$ as given in Table \ref{tab:PSRBH}. Hence, the range of $\lambda$ for which the pulsar is visible is $184.51^{\circ} - \alpha \geq \lambda \geq 175.49^{\circ} - \alpha$ and for our chosen value of $\alpha = 50^{\circ}$, the range becomes $134.51^{\circ}  \geq \lambda \geq 125.49^{\circ} $. Our choice of $\lambda=130^{\circ}$ lies in the middle of this range. This choice of $\lambda$ makes $\zeta_L = \alpha$. As we have discussed earlier, in such a case, the LoS cuts the beam through the midpoint. This fact can be seen in Figs. \ref{fig:non_distorted_beam}, \ref{fig:distorted_beam}, and \ref{fig:distorted_beam2} of sec. \ref{subsec:numerical_distort}. 

Note that, for a pulsar$-$black hole binary, we can not choose the value of $\zeta_L$ significantly different than that of $\alpha$ satisfying $ | \alpha -  \rho_{1.4, 10}^{\rm out}| \leq |\zeta_L| \leq | \alpha +  \rho_{1.4, 10}^{\rm out}|$, as the value of $ \rho_{1.4, 10}^{\rm out}$ can not be much larger than $5^{\circ}$ for normal pulsars as predicted by eq. (\ref{eq:halfopeningangleoutcone}). For a millisecond pulsar, $ \rho_{1.4, 10}^{\rm out}$ would be significantly larger and the value of $\zeta_L$ can be different than that of $\alpha$. Although the normal evolutionary scenario does not expect the pulsar in a pulsar$-$black hole binary to be a millisecond pulsar, it is possible for a dynamically formed binary in dense stellar environments like globular clusters. We do not study dynamically formed pulsar$-$black hole binaries in the present paper.

\begin{table}
\caption{Parameters for a canonical pulsar$-$black hole binary. These are (from top to bottom), the mass of the pulsar ($M_p$), the mass of the black hole ($M_c$), the spin period of the pulsar ($P_{\rm s}$), the rate of change of the spin period of the pulsar ($\dot{P}_{\rm s}$), the orbital period of the binary ($P_{\rm b}$), the eccentricity ($e$), the longitude of the periastron ($\omega$), the angle between the spin axis and the magnetic axis $(\alpha$), the initial value of the angle between the ${\rm X_s}$-axis and the projection of the spin axis in the sky plane ($\eta$), and the initial value of the angle between the ${\rm Z_s}$-axis and the spin axis ($\lambda$). The values of $e, ~M_c, ~M_p, ~P_{\rm s}, ~ \dot{P}_{\rm s}$, and $P_b$ are taken from \citet{css21}. The values of other parameters are chosen arbitrarily, but satisfying the visibility condition $|180^{\circ} + \rho_{1.4, 10}^{\rm out} - \alpha| \geq |\lambda| \geq |180^{\circ} - \rho_{1.4, 10}^{\rm out} - \alpha| $. }
\label{tab:PSRBH}
\begin{tabular}{ll}
\hline
\hline
 Parameters   & Value    \\
 \hline 
 $M_p$ [${\rm M_{ \odot}}$]  & 1.8\\
 $M_c$ [${\rm M_{\odot}}$]    & 14.5 \\
  $P_{\rm s}$ [s] & 1.67\\
 $\dot{P}_{\rm s}$ [${\rm s \, s}^{-1}$] & $1.38\times 10^{-15}$\\
 $P_{\rm b}$ [days] & 5.5\\
 $e$ &0.35\\ 
 $\omega$ [deg] &  73.804\\
   $\alpha$ [deg] & 50 \\
   $\eta$ [deg] & 45 \\
   $\lambda$ [deg] & 130 \\
\hline
\end{tabular}
\end{table}

\begin{figure*}
\begin{subfigure}[b]{0.42\textwidth}
\includegraphics[width=\textwidth]{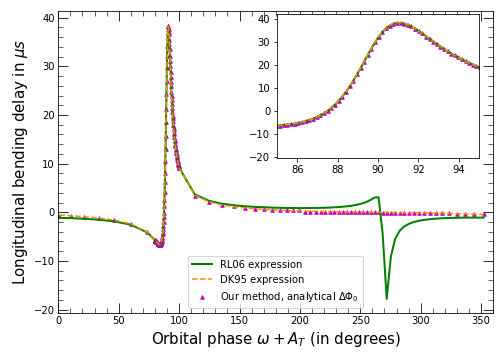}
	\label{fig:Longitudinaldelaycomparison_i87_5}
\end{subfigure}
\begin{subfigure}[b]{0.42\textwidth}
\includegraphics[width=\textwidth]{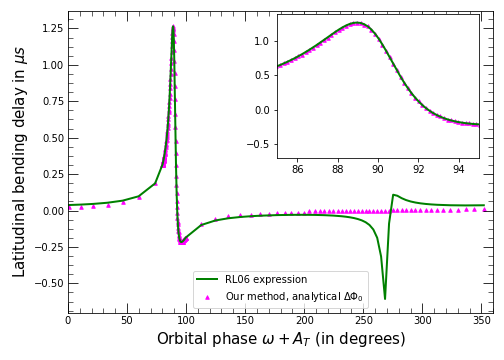}
		\label{fig:Latitudinaldelaycomparison_i87_5}
\end{subfigure}
\caption{Comparison of the values of the bending delays estimated using our method and the expressions of \citetalias{Rafikov2005} over one full orbit for a hypothetical pulsar$-$black hole binary. Here we choose $i=87.5^\circ$, and all other parameters are the same as those given in Table \ref{tab:PSRBH}. The left panel shows the longitudinal bending delay and the right panel shows the latitudinal bending delay. For the case of the longitudinal bending delay, we use the expression given by  \citetalias{dk1995} too. While computing the values of the latitudinal bending delays in our method, we use the value of $\Delta \Phi_0$ obtained using eqs. (\ref{eq:pulse_width}) and (\ref{eq:delphi}).}
\label{fig:comparison_i_89}
\end{figure*}

In Fig. \ref{fig:comparison_i_89}, we plot the bending delays in microseconds along the abscissa and the orbital phase in degrees along the ordinate over a full orbit for such a pulsar$-$black hole binary and compare the values of the bending delays estimated using our formalism (magenta triangles) with those estimated using the expressions given by \citetalias{Rafikov2005} (green lines) and \citetalias{dk1995} (orange lines) with $i=87.5^\circ$ and the values of all other parameters are the same as those given in Table \ref{tab:PSRBH}. Here, we used eqs. (\ref{eq:pulse_width}) and (\ref{eq:delphi}) to calculate the value of $\Delta \Phi_0$. We see that the values of the bending delays obtained in our method match with those obtained using the expressions given by \citetalias{Rafikov2005} except near the orbital phase of $270^{\circ}$. This is an artifact of the expressions of \citetalias{Rafikov2005}, which are valid only near the orbital phase of $90^\circ$, and hence should not be used for the values of the orbital phases away from $90^{\circ}$. On the other hand, we see that the values of the longitudinal bending delays obtained in our method match with those obtained using the expressions given by \citetalias{dk1995} throughout the orbit. However, \citetalias{dk1995} did not provide any expression for the latitudinal bending delay.

\begin{figure*}
\begin{subfigure}[b]{0.40\textwidth}
\includegraphics[width=\textwidth]{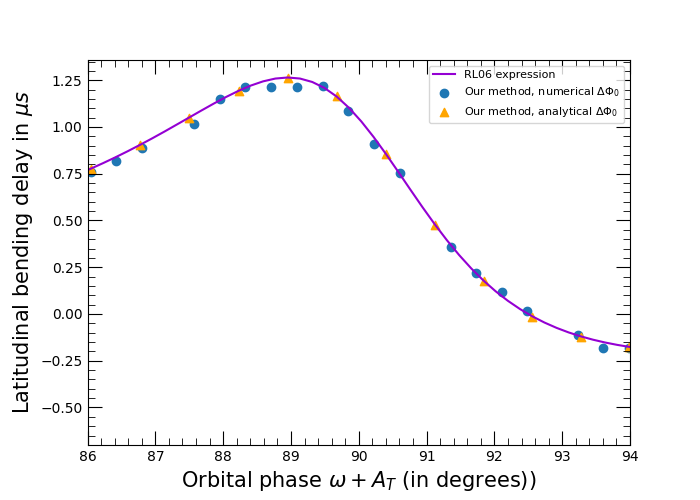}
\caption{$i=87.5^{\circ}$}
	\label{fig:LatDelayThreemethod_i87_5}
\end{subfigure}
\begin{subfigure}[b]{0.40\textwidth}
\includegraphics[width=\textwidth]{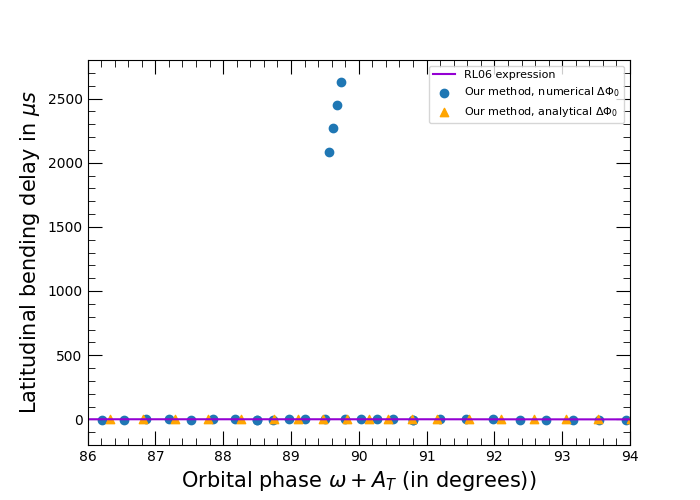}
\caption{$i=88.0^{\circ}$}
	\label{fig:LatDelayThreemethod_i88}
\end{subfigure}
\begin{subfigure}[b]{0.40\textwidth}
\includegraphics[width=\textwidth]{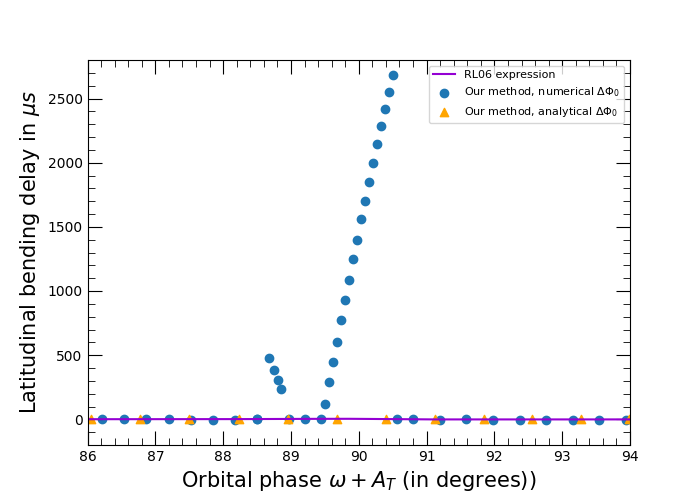}
\caption{$i=89.3^{\circ}$}
		\label{fig:LatDelayThreemethod_i89pt3}
\end{subfigure}
\begin{subfigure}[b]{0.40\textwidth}
\includegraphics[width=\textwidth]{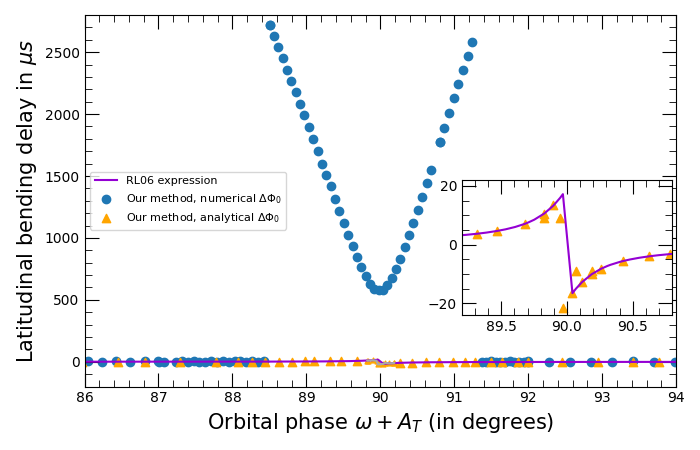}
\caption{$i=90^{\circ}$}
		\label{fig:LatDelayThreemethod_i90}
\end{subfigure}
\caption{Comparison of the values of the latitudinal bending delay estimated by three different methods for different values of $i$ for a hypothetical pulsar$-$black hole binary. The values of all other parameters are the same as those given in Table \ref{tab:PSRBH}.}
\label{fig:LatDelayThreemethod}
\end{figure*}

\begin{figure*}
\begin{subfigure}[b]{0.42\textwidth}
	\includegraphics[width=\textwidth]{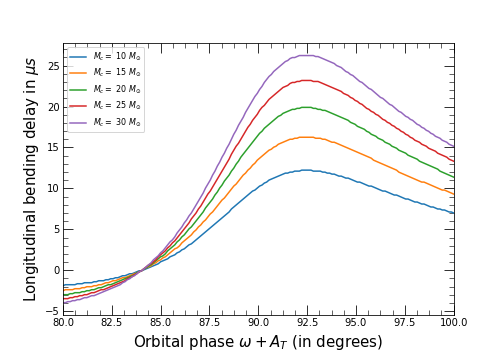}
	\caption{Longitudinal delay curves for different values of $M_{\rm c}$.}
	\label{long_diff_mass}
\end{subfigure}
\begin{subfigure}[b]{0.42\textwidth}
	\includegraphics[width=1.0\textwidth]{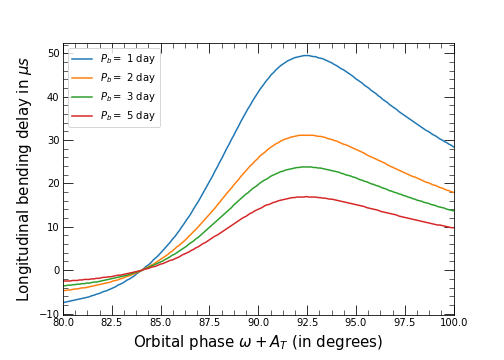}
	\caption{Longitudinal delay curves for different values of $P_{\rm b}$.}
	\label{long_diff_period}
\end{subfigure}
\begin{subfigure}[b]{0.42\textwidth}
	\includegraphics[width=1.0\textwidth]{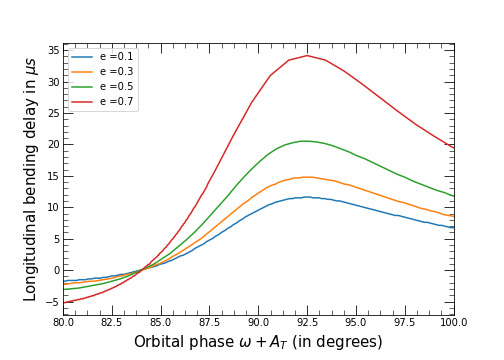}
	\caption{Longitudinal delay curves for different values of $e$.}
	\label{long_diff_eccentricity}
\end{subfigure}
\begin{subfigure}[b]{0.42\textwidth}
	\includegraphics[width=1.0\textwidth]{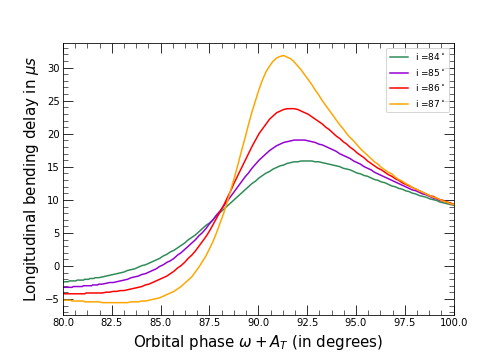}
	\caption{Longitudinal delay curves for different values of $i$.}
	\label{long_diff_inclination}
\end{subfigure}
\begin{subfigure}[b]{0.42\textwidth}
	\includegraphics[width=1.0\textwidth]{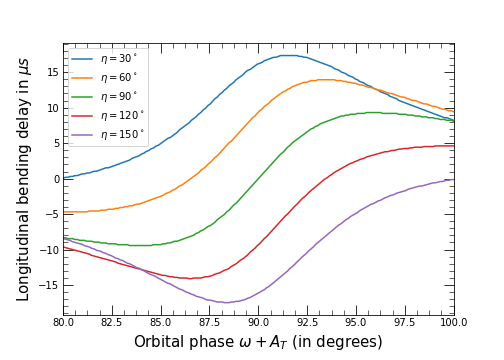}
	\caption{Longitudinal delay curves for different positive values of $\eta$.}
	\label{long_diff_eta_positive}
\end{subfigure}
\begin{subfigure}[b]{0.42\textwidth}
	\includegraphics[width=1.0\textwidth]{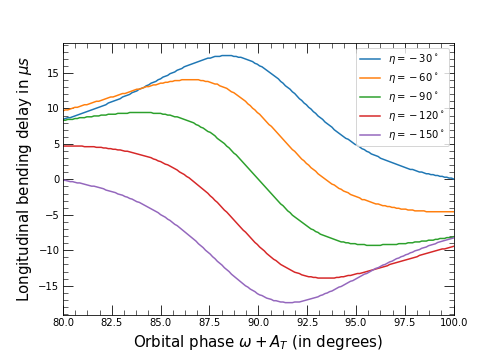}
	\caption{Longitudinal delay curves for different negative values of $\eta$.}
	\label{long_diff_eta_negative}
\end{subfigure}
\caption{The longitudinal bending delay curves for hypothetical pulsar$-$black hole binaries, with changing one parameter in each panel and keeping all the other parameters the same as those mentioned in Table \ref{tab:PSRBH}. We use $i=84^{\circ}$ except in panel (d) where we vary $i$. }
\label{longitudinal_bending_delay_diff_parameters}
\end{figure*}

\begin{figure*}
\begin{subfigure}[b]{0.42\textwidth}
	\includegraphics[width=\textwidth]{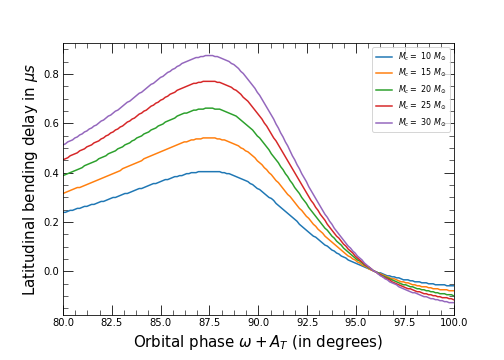}
	\caption{Latitudinal delay curves for different values of $M_{\rm c}$.}
	\label{lati_diff_mass}
\end{subfigure}
\begin{subfigure}[b]{0.42\textwidth}
	\includegraphics[width=1.0\textwidth]{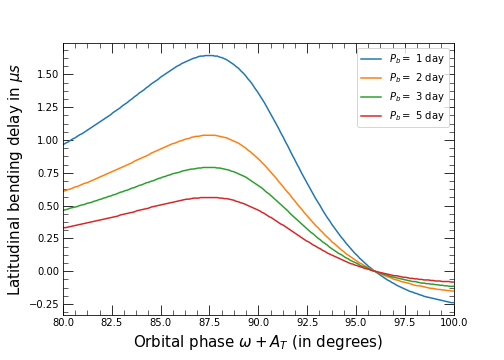}
	\caption{Latitudinal delay curves for different values of the $P_{\rm b}$.}
	\label{lati_diff_period}
\end{subfigure}
\begin{subfigure}[b]{0.42\textwidth}
	\includegraphics[width=1.0\textwidth]{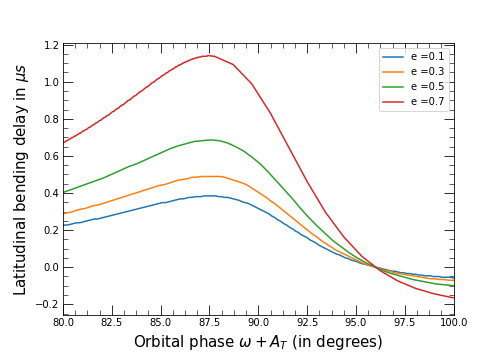}
	\caption{Latitudinal delay curves for different values of $e$.}
	\label{lati_diff_eccentricity}
\end{subfigure}
\begin{subfigure}[b]{0.42\textwidth}
	\includegraphics[width=1.0\textwidth]{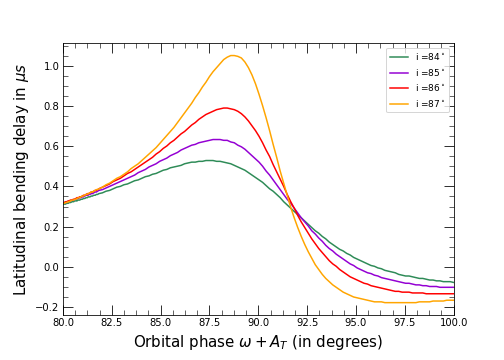}
	\caption{Latitudinal delay curves for different values of $i$.}
	\label{lati_diff_inclination}
\end{subfigure}
\begin{subfigure}[b]{0.42\textwidth}
	\includegraphics[width=1.0\textwidth]{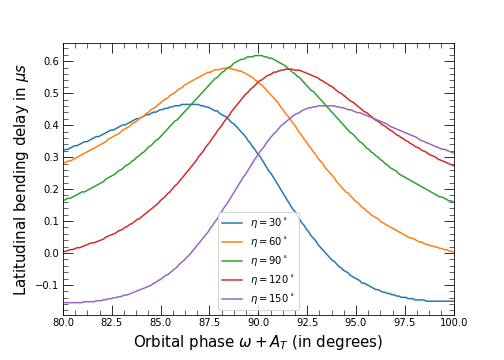}
	\caption{Latitudinal delay curves for different positive values of $\eta$.}
	\label{lati_diff_eta_positive}
\end{subfigure}
\begin{subfigure}[b]{0.42\textwidth}
	\includegraphics[width=1.0\textwidth]{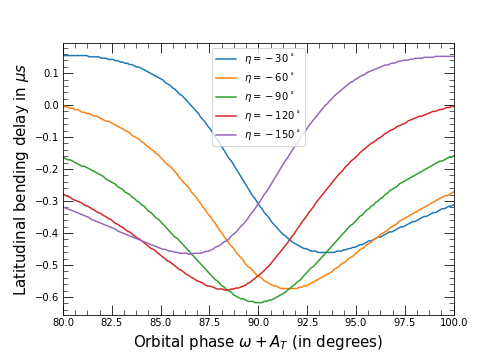}
	\caption{Latitudinal delay curves for different negative values of $\eta$.}
	\label{lati_diff_eta_negative}
\end{subfigure}
\caption{The latitudinal bending delay curves for hypothetical pulsar$-$black hole binaries, with changing one parameter in each panel and keeping all the other parameters the same as those mentioned in Table \ref{tab:PSRBH}. We use $i=84^{\circ}$ except in panel (d) where we vary $i$. }
\label{latitudinal_bending_delay_diff_parameters}
\end{figure*}

Next, we explore whether the values of the latitudinal bending delays would remain the same if we calculate $\Delta \Phi_0$ numerically. In Fig. \ref{fig:LatDelayThreemethod}, we compare the values of the latitudinal bending delay calculated using the two methods of calculating $\Delta \Phi_0$ as well as using the expression of \citetalias{Rafikov2005} for different values of $i$. We see that all three methods agree unless $i \geq 88^{\circ}$. Even for the case of $i \geq 88^{\circ}$, the mismatch between the two methods of calculating $\Delta \Phi_0$ is significant only near the orbital phase of $90^{\circ}$. The higher is the value of the $i$, the wider is the range of the orbital phase where the two methods of calculating $\Delta \Phi_0$ give different results. Where the two methods differ, the value of $\Delta \Phi_0$ obtained numerically is much larger than the value predicted by eqs. (\ref{eq:pulse_width}) and (\ref{eq:delphi}), resulting a larger value of the latitudinal bending delay. This happens because of the increased size of the beam due to the bending as seen in Figs. \ref{fig:distorted_beam} and \ref{fig:distorted_beam2}. The results obtained in our method with the value of $\Delta \Phi_0$ estimated using eqs. (\ref{eq:pulse_width}) and (\ref{eq:delphi}) disagrees with the expressions of \citetalias{Rafikov2005} only when the values of both of the inclination angle and the orbital phase are very close to $90^{\circ}$ as seen in the zoomed plot of Fig. \ref{fig:LatDelayThreemethod_i90}. This was the case for the double pulsar too. As calculating $\Delta \Phi_0$ numerically is computationally very costly, it is preferable to calculate $\Delta \Phi_0$ analytically using eqs. (\ref{eq:pulse_width}) and (\ref{eq:delphi}). As we have seen, for a pulsar$-$black hole binary with parameters the same as those given in Table \ref{tab:PSRBH}, the analytical values of $\Delta \Phi_0$ are correct when $i \leq 87.5^{\circ}$. In Figs. \ref{longitudinal_bending_delay_diff_parameters} and \ref{latitudinal_bending_delay_diff_parameters}, where we vary parameters, we use analytical values of $\Delta \Phi_0$ and $i = 84^{\circ}$ as when we get the maximum latitudinal bending for $P_b=1$ day, we found that the numerical and analytical values of the latitudinal bending delays match at $i = 84^{\circ}$ or smaller. 

In Fig. \ref{fig:LatDelayThreemethod_i90}, we also see that when $i=90^{\circ}$, the values of the latitudinal bending delay calculated using the numerically computed values of $\Delta \Phi_0$ are larger at the values of the orbital phase near $90^{\circ}$ than that exactly at $90^{\circ}$. This can be understood from Fig. \ref{fig:distorted_beam2} where $i=90^{\circ}$. Fig. \ref{fig:bending_at_phase_89} shows the structure of the beam with bending for $\omega+A_T=89^{\circ}$. In the ninth panel this figure, the LoS (denoted by a magenta diamond) exited the theoretical boundary of the beam depicted by the red circle, but due to the bending, it is still surrounded by light rays and hence inside the distorted beam. Similarly, Fig. \ref{fig:bending_at_phase_91} shows the structure of the beam with bending for $\omega+A_T=91^{\circ}$. In the first panel this figure, the LoS has not yet entered the theoretical boundary of the beam denoted by the red circle, but due to the bending, it is still surrounded by light rays and hence inside the distorted beam. These facts also lead to a tail on the right side of the pulse profile for $\omega+A_T=89^{\circ}$ and $i=90^{\circ}$ (Fig. \ref{pulse_orbitalphase89_i_90}) and on the left side of the pulse profile for $\omega+A_T=91^{\circ}$ and $i=90^{\circ}$ (Fig. \ref{pulse_orbitalphase91_i_90}). These tails are of very low intensity, about 1 to 5 in our units of intensity. On the other hand, in Fig.\ref{fig:bending_at_i_90}, where the values of both of the inclination angle and orbital phase are $90^{\circ}$, there is no light ray around the LoS when it is outside the red circle (the first and the ninth panels). However, in this case there are some light rays near around the LoS when it is on the theoretical boundary (the second and the eighth panel). These also lead to large values of the latitudinal bending delay, although smaller than those at the orbital phase of $89^{\circ}$ and $91^{\circ}$ (all three are for $i=90^{\circ}$). That is why the pulse profile with bending starts before and ends after the pulse profile without bending for these set of parameters, see Fig. \ref{pulse_phase_90_i_90}. More details about the beams shown in Figs. \ref{fig:distorted_beam} and \ref{fig:distorted_beam2} and pulse profiles shown in Figs. \ref{pulse_orbital_phase_90} and \ref{pulse_difforbital_phase_i_90} will be discussed in the next section.

In Fig. \ref{fig:LatDelayThreemethod_i89pt3}, when $i=89.3^{\circ}$, we see that there is a small range of the orbital phase less than $89^{\circ}$ where there are mild disagreements between the results obtained by the two methods of estimating $\Delta \Phi_0$ whereas there is a comparatively wider range of the orbital phase greater than $89.5^{\circ}$ where there are strong disagreements between the results obtained by the two methods of estimating $\Delta \Phi_0$. In Fig. \ref{fig:LatDelayThreemethod_i88}, where $i=88^{\circ}$, we see the disagreement between the two methods of computing $\Delta \Phi_0$ only when the orbital phase is greater than $89.5^{\circ}$, and the disagreement is always large. This happens because for $i=88^{\circ}$, for a range of orbital period larger than $89.5^{\circ}$, the LoS gets surrounded by the light rays before its entry to the theoretical boundary, whereas when the orbital phase is less than $89^{\circ}$, the LoS does not get surrounded by the light rays at all when it is outside the theoretical boundary of the beam. Similarly, for $i=89.3^{\circ}$, for a range of orbital period larger than $89.5^{\circ}$, the LoS gets surrounded by the light rays before its entry to the theoretical boundary, and for a comparatively smaller range of orbital period less than $89^{\circ}$, the LoS gets surrounded by the light rays after its exit to the theoretical boundary.

We now explore the dependence of the longitudinal and the latitudinal bending delays in our formalism on various parameters, e.g., $M_c$, $P_{\rm b}$, $e$, $i$ and $\eta$, one at a time, in Figs. \ref{longitudinal_bending_delay_diff_parameters} and \ref{latitudinal_bending_delay_diff_parameters}, respectively. In both of the cases, when we vary other parameters, we keep the value of $i$ fixed at $84^{\circ}$. When we vary $i$, we keep its value lower than $87.5^{\circ}$. The reason for making such choices for the value of $i$ is the fact that among all the combinations of the values of various parameters, the maximum bending is obtained for $P_{b}=1$ day. In this case, we have checked that the the value of $\Delta \Phi_0$ calculated using eqs. (\ref{eq:pulse_width}) and (\ref{eq:delphi})match  with the value obtained numerically if $i \leq 84^{\circ}$. However, when we vary $i$ and keep all other parameters the same as those given in Table \ref{tab:PSRBH}, such agreement can be obtained upto $i=87.5^{\circ}$. As theoretically allowed values of the mass of a neutron star lie in the range of $\sim 1 - 3~{\rm M_{\odot}}$ and the variation of $M_p$ in this small range would not affect the value of the bending delays, we keep the value of $M_p$ fixed as that mentioned in Table \ref{tab:PSRBH}. Note that, in all of the plots, instead of full orbits, we have chosen the range of the orbital phase (along the abscissa) between $80^{\circ} - 100^{\circ}$ to have the maximum delay around the middle of the plots.

Figs. \ref{long_diff_mass} and \ref{lati_diff_mass} show the values of the longitudinal and the latitudinal bending delays, respectively, for the values of $M_{\rm c}$ as 10, 15, 20, 25, and 30 solar masses. As expected, the bending delays are larger for higher values of $M_{\rm c}$. Figs. \ref{long_diff_period} and \ref{lati_diff_period} show the values of the longitudinal and the latitudinal bending delays, respectively, for the values of $P_{\rm b}$ as 1, 2, 3, and 5 days. The bending effect becomes milder for larger values of $P_{\rm b}$ as those imply wider orbits since we have kept the masses of the objects fixed here. Figs. \ref{long_diff_eccentricity} and \ref{lati_diff_eccentricity} show the values of the longitudinal and the latitudinal bending delays, respectively, for the values of $e$ as 0.1, 0.3, 0.5, 0.7. We see that the maximum values of the bending delays are larger for higher values of the eccentricity. This happens because for a higher eccentricity, the same size of the orbit (i.e., the same values of $P_{\rm b}$, $M_p$, and $M_c$) has a shorter distance between the pulsar and the companion near the periastron and nearby orbital phases. For our choice of $\omega=73.804^{\circ}$, the orbital phase corresponding to the maximum delay ($\omega + A_T  \simeq 90^{\circ}$) is close to the periastron, i.e., at $A_T \simeq 16.2^{\circ}$. For a different choice of $\omega$ giving the maximum delay closer to the apastron, one would see that the maximum value of the bending delays are larger for lower eccentricities. This would happen because for the orbital phases closer to the apastron, a more eccentric orbit has a larger distance between the pulsar and the companion than that of a less eccentric orbit of the same size.

Figs. \ref{long_diff_inclination} and \ref{lati_diff_inclination} show the values of the longitudinal and the latitudinal bending delays respectively for the values of $i$ as 85$^{\circ}$, 86$^{\circ}$, and 87$^{\circ}$. As expected, the bending effect becomes stronger when the value of $i$ increases. In Figs. \ref{long_diff_eta_positive}, \ref{long_diff_eta_negative}, \ref{lati_diff_eta_positive}, and \ref{lati_diff_eta_negative} we study the effect of the value of $\eta$ on the value of the bending delays. For both of the longitudinal and the latitudinal delays, we use both positive and negative values of $\eta$. We see that the magnitude of the maximum value of the longitudinal bending delay increases when $| \eta |$ decreases while that of the latitudinal bending delay decreases when $| \eta |$ decreases. The sign reversal of $\eta$ flips the shape of the longitudinal delay curve horizontally and the latitudinal delay curve vertically.

From different sub-panels of Figs. \ref{longitudinal_bending_delay_diff_parameters} and \ref{latitudinal_bending_delay_diff_parameters}, we also see that the value of the orbital phase corresponding to the strongest bending depends only on the values of $\eta$ and $i$. However, this value is always close to $90^{\circ}$. In these figures, we do not show the dependence of the bending delays on the value of $\lambda$, as already discussed, the pulsar will not be visible at all for a significantly different value of $\lambda$. 

\subsection{Distortion of the beam and the change in the pulse profile due to the bending}
\label{subsec:numerical_distort}
We first calculate the emission height $h_{1.4, {\rm em}}$ at $\nu=1.4$ GHz using eq. (\ref{eq:emheight}) and the values of ${P_{\rm s}}$ and $\dot{P}_{\rm s}$ given in Table \ref{tab:PSRBH}. The ${\rm X_m Y_m}$ plane, on which we create the intensity distributions, is defined at $h_{1.4, {\rm em}}$. Then, using eq. (\ref{eq:halfopeningangleToRadius}), we calculate the values of $\mathcal{R}_{\rm 1.4, peak}^{\rm core}=0$, $\mathcal{R}_{\rm 1.4, peak}^{\rm in}$, $\mathcal{R}_{\rm 1.4, peak}^{\rm out}$, $\mathcal{R}_{\rm 1.4, 50}^{\rm core}$, $\mathcal{R}_{\rm 1.4, 10}^{\rm in}$, and $\mathcal{R}_{\rm 1.4, 10}^{\rm out}$ with the help of the values of the corresponding half-opening angles obtained from eqs. (\ref{eq:halfopeningangle}) and $h_{1.4, {\rm em}}$. Here $\mathcal{R}_{\rm 1.4, peak}^{\rm core}$, $\mathcal{R}_{\rm 1.4, peak}^{\rm in}$, and $\mathcal{R}_{\rm 1.4, peak}^{\rm out}$ are the radii of the beam corresponding to the peak of core, the inner cone, and the outer cone, respectively. Similarly, $\mathcal{R}_{1.4, 50}^{\rm core}$ is the radius of the beam corresponding to the $50\%$ of the maximum intensity of the core, $\mathcal{R}_{1.4, 10}^{\rm in}$ and $\mathcal{R}_{1.4, 10}^{\rm out}$ are the radii of the beam corresponding to the $10\%$ of the maximum intensities for the inner and the outer cones, respectively. 

Now, for each component of the beam, we select the mean and the standard deviation of the Gaussian intensity distribution (see eq. (\ref{eq:gauss})) in such a way that $\mu_{\rm core}=\mathcal{R}_{\rm 1.4, peak}^{\rm core}=0$, $\mu_{\rm in}=\mathcal{R}_{\rm 1.4, peak}^{\rm in}$ and $\mu_{\rm out}=\mathcal{R}_{\rm 1.4, peak}^{\rm out}$. Here $\mu_{\rm core}$, $\mu_{\rm in}$, and $\mu_{\rm out}$ represent the mean of the intensity distributions of the core, the inner cone, and the outer cone, respectively. The standard deviations of the intensity distributions of these components are denoted by $\sigma_{\rm core}$, $\sigma_{\rm in}$, and $\sigma_{\rm out}$, respectively. We choose the values of $\sigma_{\rm core}=6.018$, $\sigma_{\rm in}=1.987$, and $\sigma_{\rm out}=3.401$ in such a way that the values of $\mathcal{R}_{1.4, 50}^{\rm core}$, $\mathcal{R}_{1.4, 10}^{\rm in}$, and $\mathcal{R}_{1.4, 10}^{\rm out}$ of the intensity distributions match with those obtained using eqs. (\ref{eq:halfopeningangle}), (\ref{eq:emheight}), and (\ref{eq:halfopeningangleToRadius}). 

The choice of the values of the multiplicative constant in eq. (\ref{eq:gauss}) gives the relative strengths of different components. In this work, we choose $I_0^{\rm core}=300 \, {\rm km^{-2}}$, $I_0^{\rm in}=500 \, {\rm km^{-2}}$, and $I_0^{\rm out}=700 \, {\rm km^{-2}}$, where $I_0^{\rm core}$, $I_0^{\rm in}$, and $I_0^{\rm out}$ denote the constant for the core, the inner cone, and the outer cone, respectively. Instead of a physical unit for the intensity, we have chosen its unit as area inverse so that the multiplication of the intensity by an area will give a dimensionless quantity that can be taken proportional to the number of the light rays in that area. This particular beam model, accompanied by the choice of $\alpha = \zeta_L$ that makes the LoS cut the beam through the midpoint of the beam, lead to five-component pulse profiles whose outer components are the strongest and the core component is the weakest. These profiles would be close to the observed profiles of PSR B1237$+$25 as reported by \citet{sgg95}. We have not modelled the scattering tail and the noise in the profiles, as getting realistic profiles is not our aim, we want to demonstrate the effect of the light bending using model profiles. 

In our model, the intensity of the core falls off very rapidly, it becomes $10\%$ of the maximum value at the radius of 12.91 km on the ${\rm X_m Y_m}$ plane, before the intensity of the inner cone becomes significant - the intensity of the inner cone becomes $10\%$ of the maximum value only at the radius of 17.6 km, then reaching the peak at the radius of 21.86 km, it again falls off to $10\%$ of the maximum value at the radius of 26.12 km. However, the intensity of the outer cone becomes significant even before this, it reaches $10\%$ of the maximum value at the radius of 19.89 km. Hence, there is a significantly large region where the intensities of both of the cone components are significantly large and it becomes difficult to identify the components. To avoid this, we truncate the inner core outwards and the outer cone inwards at the radius of 25.22 km where the intensities of these two components are equal.

After getting the intensity distributions for all three components of the beam as described above, we generate a number of starting points of the light rays on the ${\rm X_m Y_m}$ plane in such a way that the surface densities of these points match the intensity distributions of the three components. To achieve this, first, near the centre of the cross-section, where only the core component is important, the number of light rays ($N_{\rm cent}$) in a small area of radius $\mathcal{R}_{\rm cent}$ around the centre is taken as the value of $I_0^{\rm core}\pi \mathcal{R}_{\rm cent}^2$, rounded off to the nearest integer. We choose the value of $ \mathcal{R}_{\rm cent}$ as 0.05 km. We assume that in this area, these $N_{\rm cent}$ light rays are uniformly distributed in $r$ and $\theta$, where $r$ and $\theta$ are the polar coordinates on the ${\rm X_m Y_m}$ plane. Outside this central region, we proceed as follows. At a radial position $r$, we calculate the intensity $I(r)$ for each of the components of the beam following the distributions discussed earlier, and choose a small annular region of width $dr= 0.01$ km around $r$. The area of this annular region is $\mathcal{A}=2\pi r dr$ and the number of light rays $N_{\rm ray}$ in that annular region is obtained by rounding off $I(r) \mathcal{A}$ to the nearest integer. We calculate the number of light rays (or more specifically, the starting point of the light rays) for each of the components separately and distribute those over $\mathcal{\theta}$ with a uniform distribution. Note that, for each value of $r$, there is always a non-zero value of $I(r)$ for all three components of the beam. However, when the value of $I(r)$ is so small that the value of $I(r) \mathcal{A}$ of a particular component is less than 0.5, we don't have any contribution from that component in the ray distribution. We also emphasise the fact that, these numbers are not the actual numbers of the light rays that the beam has in reality. The surface density of the light rays generated in this way just corresponds to the intensity distribution on the cross-section of the beam as per the model we consider.

\begin{figure}
	\includegraphics[width=90mm]{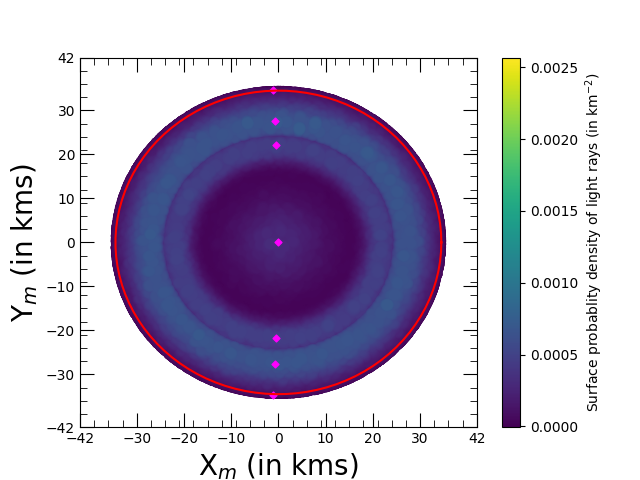}
	\caption{The surface probability density of the light rays on the cross section of the beam of a pulsar on the ${\rm X_m Y_m}$ plane if the bending was absent. The red circle represents is the circle of radius $\mathcal{R}_{\rm 1.4, 10}^{\rm out}$. The magenta diamonds ($\diamond$) show the position of the LoS across the beam at different times. The middle point is at a time $t_{\rm mid}$ when the LoS is at the centre of the beam. The three points above the middle point are for $t_{\rm mid} - 0.0274 ~\rm{s}$, $t_{\rm mid} - 0.0218 ~\rm{s}$, and $t_{\rm mid} - 0.0174 ~\rm{s}$ (from top to bottom). The three points below the middle points are for  $t_{\rm mid} + 0.0174  ~\rm{s}$, $t_{\rm mid} + 0.0218 ~\rm{s}$, and $t_{\rm mid} + 0.0274 ~\rm{s}$ (from top to bottom).}
	\label{fig:non_distorted_beam}
\end{figure}

To visualise the intensity distribution on the cross-section of the beam, instead of using the values of $I(r)$ obtained from eq. (\ref{eq:gauss}), we use the surface probability density of the starting points of the light rays. This is because, after bending, the intensity distribution on the cross section of the beam would be different, which we do not know a priori. To get the surface probability density of the starting points of the light rays, we first convert the polar coordinates $r, \theta$ of each of these points to Cartesian coordinates on the ${\rm X_m Y_m}$ plane. Then, we calculate the normalised surface density or the surface probability density of the light rays in this plane. To do this, we use a two-dimensional histogram with 100 bins in each dimension and calculate the probability density of the light rays in each bin as $({\rm number \ of \ rays \ in \ the \ bin }) /({\rm total\ number\ of\ rays \ \times\ bin\ area})$. In Fig. \ref{fig:non_distorted_beam}, the colour map represent the surface probability density obtained in this manner and the abscissa and the ordinate correspond to the values of the Cartesian coordinates in the ${\rm X_m Y_m}$ plane. Note that, we use the same method to get the surface probability density of apparent starting points of the light rays across the beam after bending, too. In Fig. \ref{fig:non_distorted_beam}, the magenta diamonds show the trajectory of the LoS across the beam and the red line is the circle of radius $\mathcal{R}_{1.4, 10}^{\rm out}$. Note that, in the schematic diagram of Fig. \ref{fig:spin_axis_frame}, we showed  $\mathcal{R}_{1.4, 10}^{\rm out}$ as $\mathcal{R}_{\rm beam}$.

\begin{figure*}
\begin{subfigure}[b]{1.0\textwidth}
	\includegraphics[width=\textwidth]{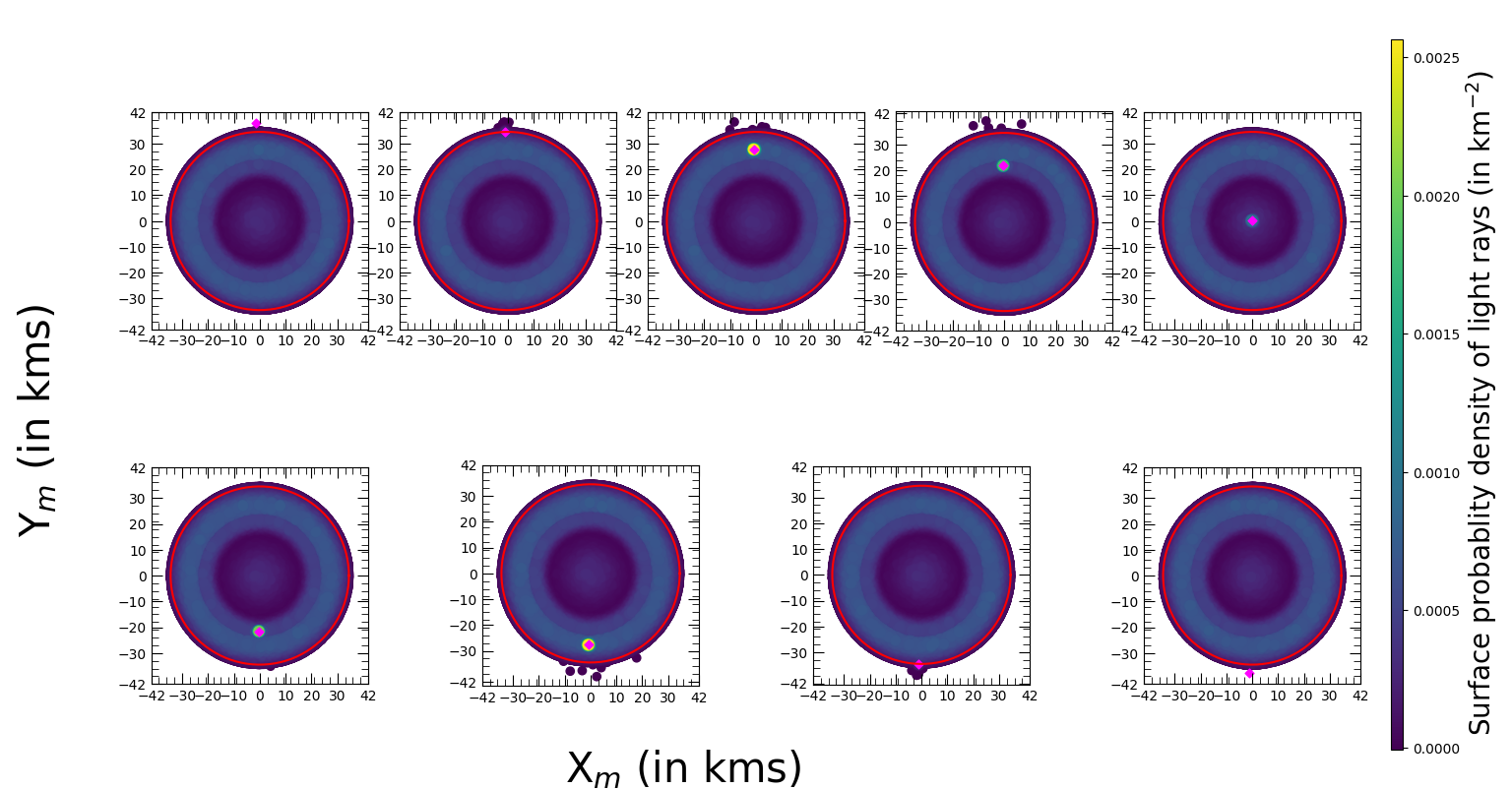}
	\caption{The surface probability density of the light rays under bending for $\omega+A_T=90^\circ$ and $i=90^\circ$. The fifth plot is at $t_{\rm mid}$, i.e., when the LoS is at the middle of the line along which it passes through the beam. The four plots before it are for  $t_{\rm mid} - 0.0298 ~\rm{s}$ (the first plot), $t_{\rm mid} - 0.0274 ~\rm{s}$ (the second plot), $t_{\rm mid} - 0.0218 ~\rm{s}$ (the third plot) and $t_{\rm mid} - 0.0174 ~\rm{s}$ (the fourth plot), while the four plots after it are for  $t_{\rm mid} + 0.0174 ~\rm{s}$ (the sixth plot), $t_{\rm mid} +  0.0218  ~\rm{s}$ (the seventh plot) $t_{\rm mid} + 0.0274 ~\rm{s}$ (the eighth plot) and $t_{\rm mid} + 0.0298 ~\rm{s}$ (the ninth plot).  }
\label{fig:bending_at_i_90}
\end{subfigure}
\begin{subfigure}[b]{1.0\textwidth}
	\includegraphics[width=\textwidth]{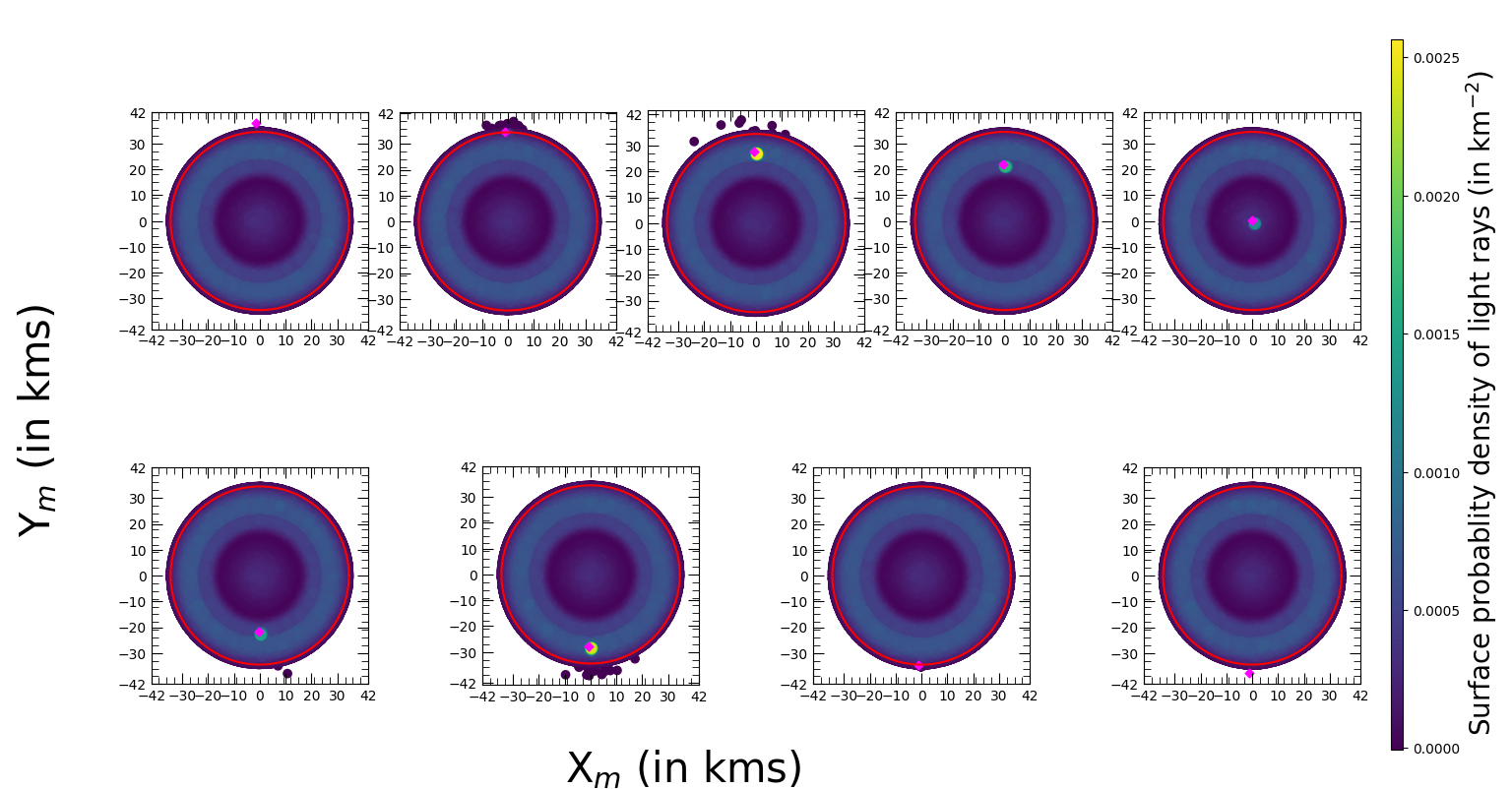}
	\caption{The surface probability density of the light rays under bending for $\omega+A_T=90^\circ$ and $i=89.8^\circ$. The fifth plot is at $t_{\rm mid}$, i.e., when the LoS is at the middle of the line along which it passes through the beam. The four plots before it are for  $t_{\rm mid} - 0.0298 ~\rm{s}$ (the first plot), $t_{\rm mid} - 0.0274 ~\rm{s}$ (the second plot), $t_{\rm mid} - 0.0218 ~\rm{s}$ (the third plot) and $t_{\rm mid} - 0.0174 ~\rm{s}$ (the fourth plot), while the four plots after it are for  $t_{\rm mid} + 0.0174 ~\rm{s}$ (the sixth plot), $t_{\rm mid} +  0.0218  ~\rm{s}$ (the seventh plot) $t_{\rm mid} + 0.0274 ~\rm{s}$ (the eighth plot) and $t_{\rm mid} + 0.0298 ~\rm{s}$ (the ninth plot). }
\label{fig:bending_at_i_89_8}
\end{subfigure} 
 \caption{The surface probability density of the light rays on the cross section of the beam on the ${\rm X_m Y_m}$ plane under the gravitational bending for a hypothetical pulsar$-$black hole binary. The top panel (panel-a) is for $i=90.0^\circ$ and the bottom panel (panel-b) is for $i=89.8^\circ$. For both of the cases, we have taken $\omega+A_T=90^\circ$ and all others parameters are taken from Table \ref{tab:PSRBH}. In each of the panels, we show the intensity distribution at nine different times. For each of these instants, in addition to the surface density of the light rays, the location where the LoS cuts the cross section has been shown with a magenta diamond ($\diamond$). In each of the plots, the red circle represents the theoretical boundary of the cross-section of the beam without bending, i.e., the circle of radius $\mathcal{R}_{\rm 1.4, 10}^{\rm out}$. } 
\label{fig:distorted_beam}
\end{figure*}

\begin{figure*}
\begin{subfigure}[b]{1.0\textwidth}
	\includegraphics[width=\textwidth]{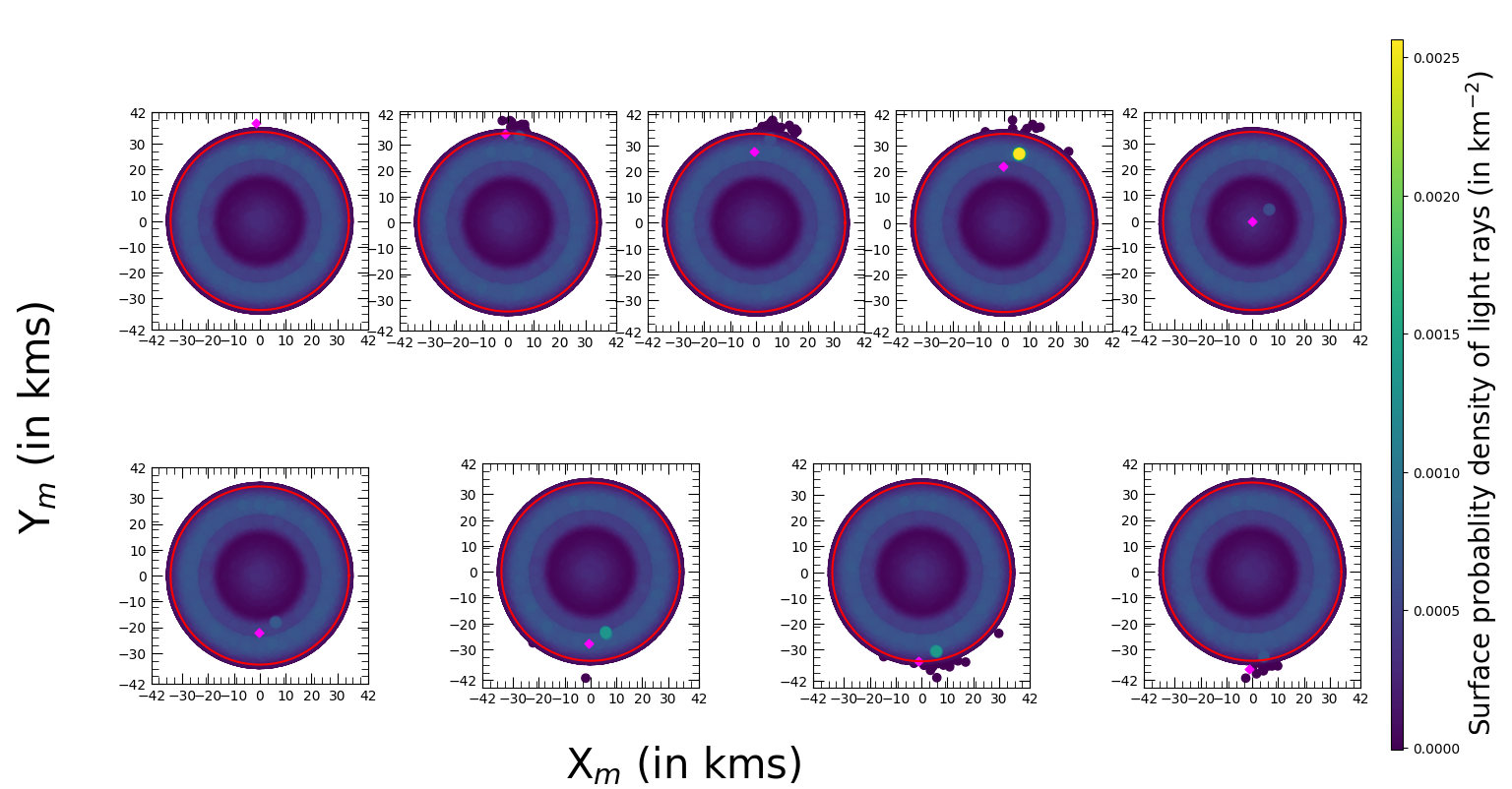}
	\caption{The surface probability density of the light rays under bending for $\omega+A_T=89^\circ$ and $i=90^\circ$. The fifth plot is at $t_{\rm mid}$, i.e., when the LoS is at the middle of the line along which it passes through the beam. The four plots before it are for  $t_{\rm mid} - 0.0298 ~\rm{s}$ (the first plot), $t_{\rm mid} - 0.0274 ~\rm{s}$ (the second plot), $t_{\rm mid} - 0.0218 ~\rm{s}$ (the third plot) and $t_{\rm mid} - 0.0174 ~\rm{s}$ (the fourth plot), while the four plots after it are for  $t_{\rm mid} + 0.0174 ~\rm{s}$ (the sixth plot), $t_{\rm mid} +  0.0218  ~\rm{s}$ (the seventh plot) $t_{\rm mid} + 0.0274 ~\rm{s}$ (the eighth plot) and $t_{\rm mid} + 0.0298 ~\rm{s}$ (the ninth plot). }
\label{fig:bending_at_phase_89}
\end{subfigure}
\begin{subfigure}[b]{1.0\textwidth}
	\includegraphics[width=\textwidth]{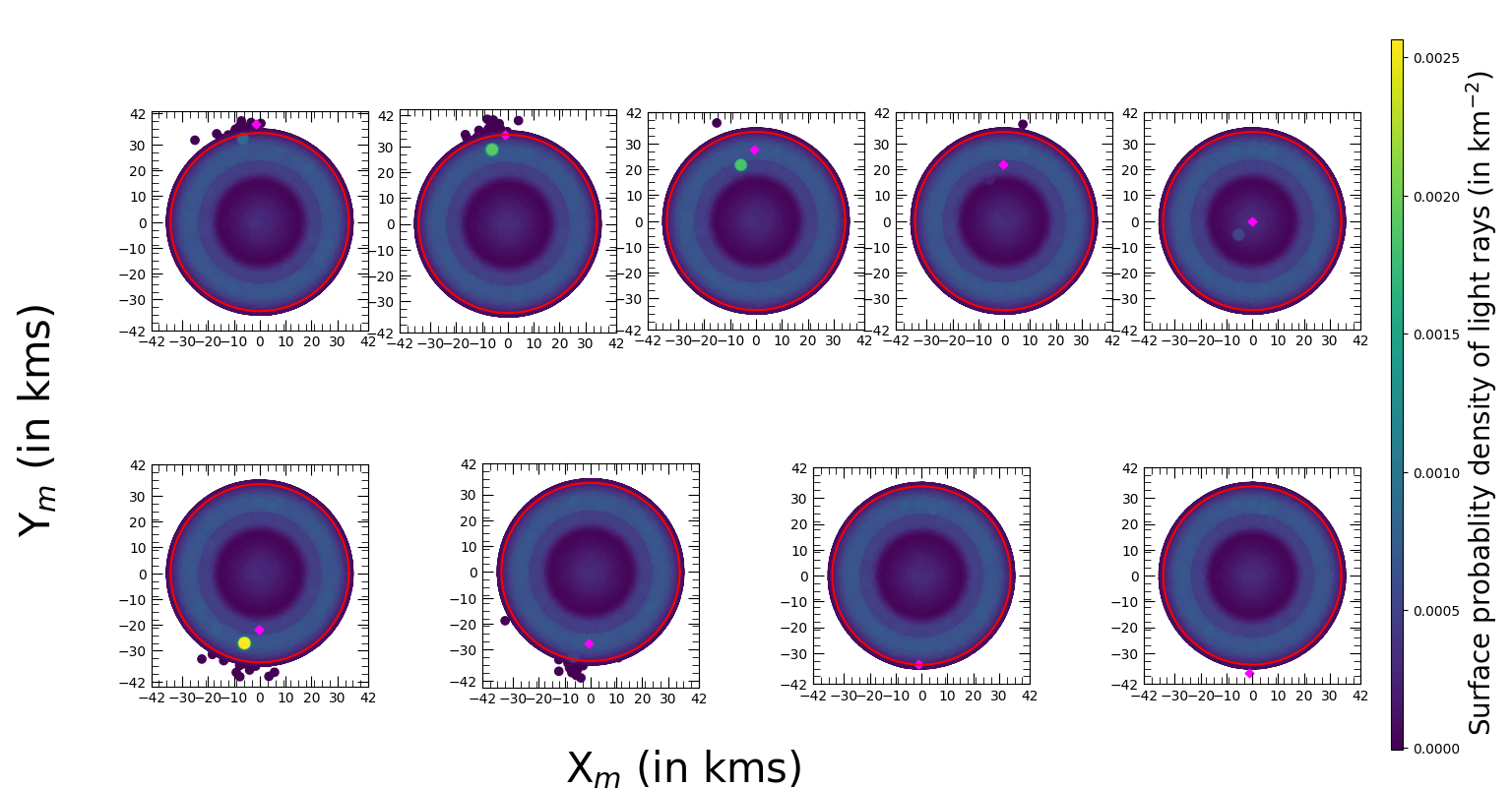}
	\caption{The surface probability density of the light rays under bending for $\omega+A_T=91^\circ$ and $i=90^\circ$. The fifth plot is at $t_{\rm mid}$, i.e., when the LoS is at the middle of the line along which it passes through the beam. The four plots before it are for  $t_{\rm mid} - 0.0298 ~\rm{s}$ (the first plot), $t_{\rm mid} - 0.0274 ~\rm{s}$ (the second plot), $t_{\rm mid} - 0.0218 ~\rm{s}$ (the third plot) and $t_{\rm mid} - 0.0174 ~\rm{s}$ (the fourth plot), while the four plots after it are for  $t_{\rm mid} + 0.0174 ~\rm{s}$ (the sixth plot), $t_{\rm mid} +  0.0218  ~\rm{s}$ (the seventh plot) $t_{\rm mid} + 0.0274 ~\rm{s}$ (the eighth plot) and $t_{\rm mid} + 0.0298 ~\rm{s}$ (the ninth plot). }
\label{fig:bending_at_phase_91}
\end{subfigure} 
 \caption{The surface probability density of the light rays on the cross section of the beam on the ${\rm X_m Y_m}$ plane under the gravitational bending for a hypothetical pulsar$-$black hole binary. The top panel (panel-a) is for $\omega+A_T=89^\circ$ and the bottom panel (panel-b) is for $\omega+A_T=91^\circ$. For both of the cases, we have chosen $i=90^\circ$ and all others parameters are taken from Table \ref{tab:PSRBH}. In each of the panels, we show the intensity distribution at nine different times. For each of these instants, in addition to the surface density of the light rays, the location where the LoS cuts the cross section has been shown with a magenta diamond ($\diamond$). In each of the plots, the red circle represents the theoretical boundary of the cross-section of the beam without bending, i.e., the circle of radius $\mathcal{R}_{\rm 1.4, 10}^{\rm out}$. } 
\label{fig:distorted_beam2}
\end{figure*}

After getting the distribution of the starting points of the light rays, we study the effect of bending on those. For this purpose, we use the parameters for the pulsar$-$black hole binary as given in Table \ref{tab:PSRBH} and apply the formalism described in sec. \ref{subsec:analytic_bendingdelay} to each of the light rays. To visualise the effect of the bending on the surface density of the light rays over the cross section of the beam well, we first choose the orbital phase as $90^\circ$. Figs. \ref{fig:bending_at_i_90} and \ref{fig:bending_at_i_89_8} show the surface probability density of the apparent starting points of the light rays after light bending for $i=90^\circ$ and $i=89.8^\circ$, respectively. In Figs. \ref{fig:bending_at_phase_89} and \ref{fig:bending_at_phase_91}, we show the surface probability density of the apparent starting point of the light rays for $i=90^\circ$ but at two different values of the orbital phase, e.g., $89^\circ$ and $91^\circ$.

In each of the cases, different panels are for different times within one rotational period of the pulsar. In one rotational period of the pulsar, although the orbital phase does not change significantly, but the direction of emission changes over time due to the rotation of the beam. This results in a change in the surface density of the light rays on the cross section of the beam after bending, as the bending of light depends on the direction of emission. In each of the plots, in addition to the surface probability density of the light rays, the location where the LoS cuts the cross section has been shown with a magenta diamond and the theoretical boundary of the cross-section of the beam without bending, i.e., the circle of radius $\mathcal{R}_{\rm 1.4, 10}^{\rm out}$ is marked with a red circle. 

When the inclination angle and the orbital phase, both are equal to $90^\circ$ (Fig. \ref{fig:bending_at_i_90}), i.e., the pulsar is at the superior conjunction, the beam emitted by the pulsar reaches the observer by crossing the companion. Although the radius of the beam is about 30 km where it starts as seen in Fig. \ref{fig:non_distorted_beam}, it becomes much larger when it crosses the companion, as obtained by using the value of $h$ as the distance between the pulsar and the companion in eq. (\ref{eq:halfopeningangleToRadius}). Hence, the companion is like a point with respect to the cross-section of the beam that cuts it. Due to its strong gravity, this point-like companion acts like a convex lens and focuses all the light rays from a ring-shaped region around it. As a result, the density of the light rays increases around the point where it cuts the companion. When we trace back the bent beam at the emission height and plot the surface probability distribution of these apparent starting points of the light rays, we see a small circular region with enhanced intensity. This feature is noticeable in various panels of Figs. \ref{fig:distorted_beam} and \ref{fig:distorted_beam2} as small circular patches with different colours that represent larger values of the surface probability density.  Note that, due to bending, sometimes there are cluster of light rays outside the circle of radius $\mathcal{R}_{\rm 1.4, 10}^{\rm out}$, and this can happen even when the LoS is outside this circle. If the LoS is surrounded by these rays, then the effective size of the beam increases, leading to large values of $\Delta \Phi_0$ when calculated numerically, and hence, large values of the latitudinal bending delay. This fact has been already discussed. In addition to those clusters, there are some light rays that seem to start far from the others. These are the rays with very low impact parameters. The exact number of such rays depends on the value of $I_0^{\rm out}$. However, these isolated points do not affect any of our calculations.

Note that, when the beam cuts the companion in the outer cone region, which is chosen to be the brightest of the components, the surface density enhancement is the largest. This is the case for the third and the seventh panels of Figs. \ref{fig:bending_at_i_90} and \ref{fig:bending_at_i_89_8}, the fourth panel of Fig. \ref{fig:bending_at_phase_89}, and the sixth panel of Fig. \ref{fig:bending_at_phase_91}. However, if the companion cuts the outer cone near its outer or inner edge, where the intensity is lower than the peak value, the  the surface density enhancement is not that large. This is the case for the seventh panel of Fig. \ref{fig:bending_at_phase_89} and the second panel of Fig. \ref{fig:bending_at_phase_91}. When the beam cuts the companion in the inner cone region, which is chosen to be fainter than the outer cone, the density enhancement is also smaller. This is the case for the fourth and the sixth panels of Figs. \ref{fig:bending_at_i_90} and \ref{fig:bending_at_i_89_8}, as well as for the sixth panel of Fig. \ref{fig:bending_at_phase_89} and for the third panel of Fig. \ref{fig:bending_at_phase_91}. When the beam cuts the companion in the core region, which is chosen to be the faintest of the components, the enhancement of the surface density is also very small. This is the case for the fifth panels of Figs. \ref{fig:bending_at_i_90}, \ref{fig:bending_at_i_89_8}, \ref{fig:bending_at_phase_89}, and \ref{fig:bending_at_phase_91}. In the fourth panel of \ref{fig:bending_at_phase_91}, the density enhancement is almost zero as the companion cuts the beam at near the outer edge of the core. In the second panels of Figs. \ref{fig:bending_at_i_90} and \ref{fig:bending_at_i_89_8}, the eighth panel of  Figs. \ref{fig:bending_at_i_90},
the second panel of Fig. \ref{fig:bending_at_phase_89} and the seventh panel of \ref{fig:bending_at_phase_91}, the beam cuts the companion at the edge of the outer cone where the surface probability distribution of the light rays is only $10\%$ of its maximum value and hence the density enhancement is not noticeable. However, distortion of the beam, i.e., cluster of light rays outside the red circle is visible. In the first and the ninth panels of Figs. \ref{fig:bending_at_i_90} and \ref{fig:bending_at_i_89_8}, as well as in the first panel of Fig. \ref{fig:bending_at_phase_89} and eighth and ninths panels of Fig. \ref{fig:bending_at_phase_91}, the companion does not cut the beam at all and we don't see any enhanced density region. We do not even see any distortion of the beam as the companion is far enough from the beam.

\begin{figure*}
\begin{subfigure}[b]{0.86\textwidth}
	\includegraphics[width=\textwidth]{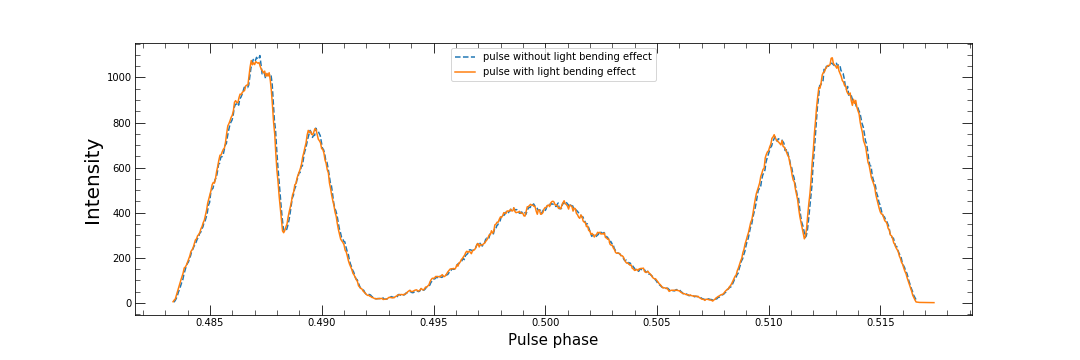}
	\caption{Pulse profiles for $\omega+A_T = 89^\circ$. }
	\label{pulse_orbitalphase89_i_90}
\end{subfigure}	
	\begin{subfigure}[b]{0.86\textwidth}
	\includegraphics[width=\textwidth]{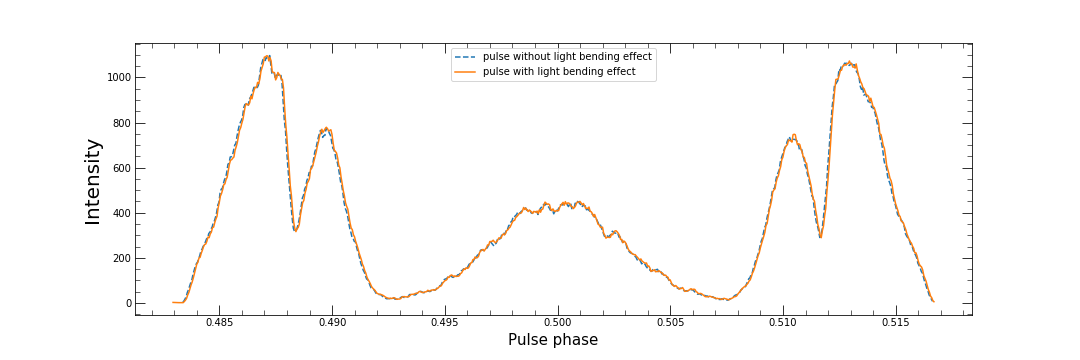}
	\caption{Pulse profiles for $\omega+A_T = 91^\circ$. }
	\label{pulse_orbitalphase91_i_90}
\end{subfigure}
\caption{Comparison between the pulse profiles with (solid orange lines) and without (blue dashed lines) the bending effect for a hypothetical pulsar$-$black hole binary with $\omega +A_T=89^\circ$ (panel a) and $\omega +A_T=91^\circ$ (panel b). In both of the cases, we have chosen $i=90^\circ$ and all other parameters are taken from Table \ref{tab:PSRBH}. Here, the intensity means the number of light rays in a circular region of radius $\mathcal{R}_{1.4, 10}^{\rm out}/50$ around the LoS. }
\label{pulse_difforbital_phase_i_90}
\end{figure*}

Another phenomenon manifests in these figures. When the pulsar is at the superior conjunction, the LoS also cuts the companion. The farther the pulsar is from the superior conjunction, the farther is the LoS from the companion. That is why in Fig. \ref{fig:bending_at_i_90}, the LoS (magenta diamonds) is always at the centre of the enhanced intensity regions while in Fig. \ref{fig:bending_at_i_89_8}, the LoS is shifted from the centre of the enhanced intensity regions, and in Figs. \ref{fig:bending_at_phase_89} and \ref{fig:bending_at_phase_91}, the LoS is even outside the enhanced intensity regions. That is why there is no difference between the pulse profiles without and with bending for $i=90^\circ$ and $\omega+A_T$ either $89^\circ$ or $91^\circ$ , except the tail at the right side for the first case (Fig.\ref{pulse_orbitalphase89_i_90}) and at the left side side for the second case (Fig. \ref{pulse_orbitalphase89_i_90}) caused by fact of the LoS being surrounded by the light rays outside the circle of radius $\mathcal{R}_{\rm 1.4, 10}^{\rm out}$ (or strong latitudinal bending), as seen in Figs. \ref{fig:bending_at_phase_89}, and \ref{fig:bending_at_phase_91}.

Next, we explore the change in the shape of the pulse profile due to the bending. To obtain a pulse profile from the distribution of the light rays on the cross section of the beam, we use a simple technique. We divide $P_{\rm s}$ into 20000 time-steps, and for each of these times, we first obtain the orientation of the beam in the I-frame (in which the LoS remain fixed) and then consider a very small circular region of radius $\mathcal{R}_{1.4, 10}^{\rm out}/50$ around the LoS on the ${\rm X_m Y_m}$ plane and check whether there is any light rays in that region. When we get a non zero value, we can say that the LoS enters into the beam. The number of the light rays inside that circular region is taken as a proxy of the intensity. Note that, this intensity is different than the surface probability density of the light rays mentioned earlier. This intensity is taken along the ordinate and the pulse phase is taken along the abscissa whenever a pulse profile is plotted. The pulse phase is defined to vary from $0 - 1$ which cover a full rotation of the magnetic axis. For the purpose of visualisation, we define the phase in such a way that the peak of the core component of the profile without the bending is at a phase 0.5. In other words, at the pulse phase of 0.5, the LoS is at the middle of the unbent beam. While plotting the profile, instead of plotting the full range of the pulse phase, we plot only the range over which the LoS is within the beam.

\begin{figure*}
\begin{subfigure}[b]{0.86\textwidth}
	\includegraphics[width=\textwidth]{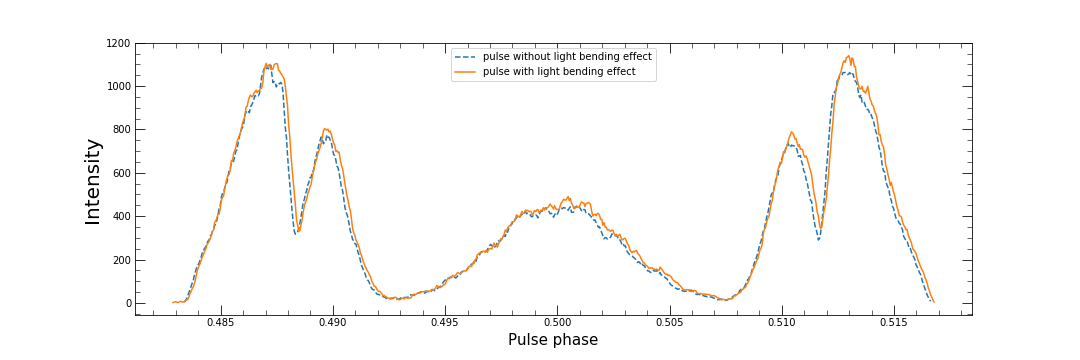}
	\caption{Pulse profiles for $i = 89.7^\circ$. }
	\label{pulse_phase_90_i_89_7}
\end{subfigure}
\begin{subfigure}[b]{0.86\textwidth}
	\includegraphics[width=\textwidth]{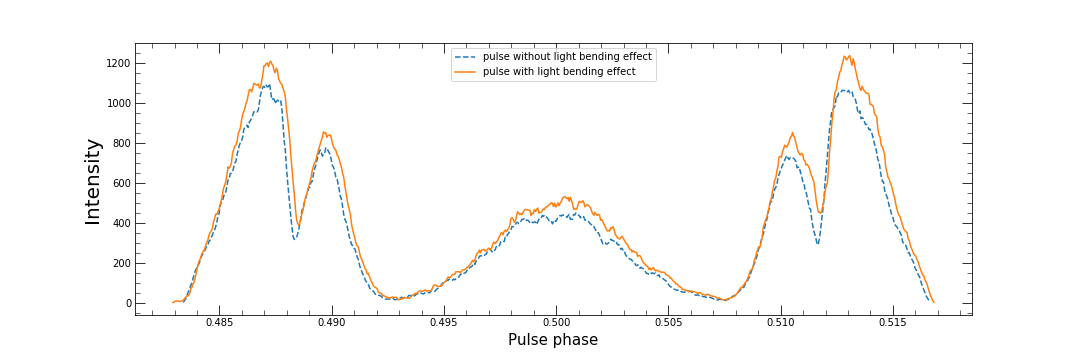}
	\caption{Pulse profiles for $i = 89.8^\circ$. }
	\label{pulse_phase_90_i_89_8}
\end{subfigure}
\begin{subfigure}[b]{0.86\textwidth}
	\includegraphics[width=\textwidth]{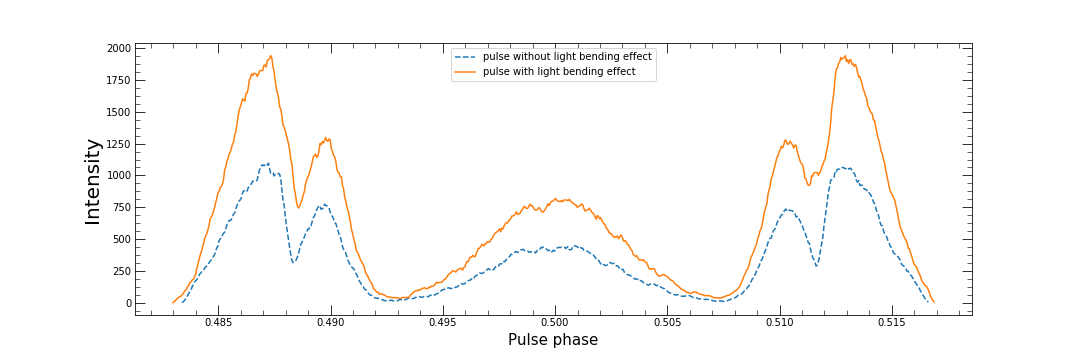}
	\caption{Pulse profiles for $i = 89.9^\circ$. }
	\label{pulse_phase_90_i_89_9}
\end{subfigure}
\begin{subfigure}[b]{0.86\textwidth}
	\includegraphics[width=\textwidth]{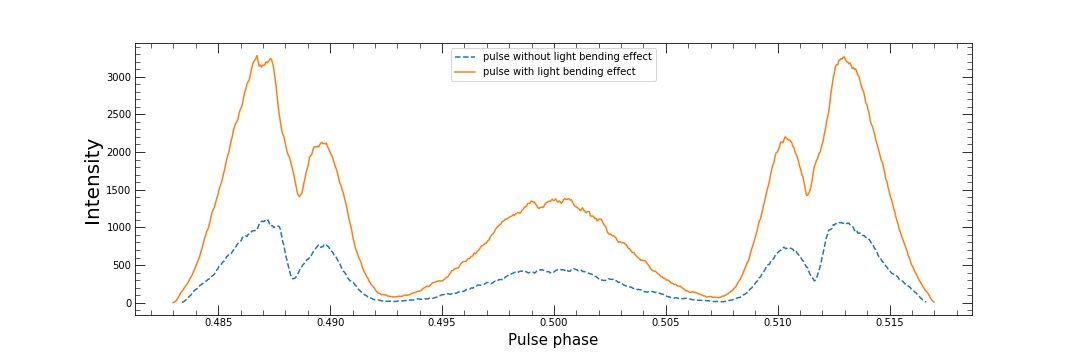}
	\caption{Pulse profiles for $i = 90^\circ$. }
	\label{pulse_phase_90_i_90}
\end{subfigure}
\caption{Comparison between the pulse profiles with (solid orange lines) and without (blue dashed lines) the bending for a hypothetical pulsar$-$black hole binary for four different inclination of the orbit. In these panels, the intensity means the number of light rays in a circular region of radius $\mathcal{R}_{1.4, 10}^{\rm out}/50$ around the LoS. For each of the cases, the pulsar is at the orbital phase $\omega +A_T=90^\circ$ and all other parameters are taken from Table \ref{tab:PSRBH}. }
\label{pulse_orbital_phase_90}
\end{figure*}

In Figs. \ref{pulse_phase_90_i_89_7} ,\ref{pulse_phase_90_i_89_8}, \ref{pulse_phase_90_i_89_9}, and \ref{pulse_phase_90_i_90}, we show the pulse profiles for the value of the orbital phase as $90^\circ$ and the values of the inclination angle as $89.7^\circ$, $89.8^\circ$, $89.9^\circ$, and $90^\circ$, respectively. In each of these cases, we compare the pulse profile under bending with the pulse profile without bending (using the distribution of the light rays of Fig. \ref{fig:non_distorted_beam}). Due to the bending, the profile strength increases as the LoS passes through the enhanced intensity region on the cross section of the beam as discussed earlier. This increase in the profile strength becomes larger as the inclination angle comes closer to $90^\circ$ due to the increase in the effects of the longitudinal bending. From the above figures, we conclude that we can get a very strong pulse when both of the inclination angle and the orbital phase are close to $90^\circ$, as the LoS comes to the centre of the enhanced intensity region (Fig. \ref{fig:bending_at_i_90}).

Note that, the profiles in Figs. \ref{pulse_orbital_phase_90} and \ref{pulse_difforbital_phase_i_90} are single pulse profiles. In reality, we work with integrated pulse profiles obtained by adding a large number of consecutive single profiles. Although not plotted, but we have checked that the nature of the integrated profiles remain the the same as the single pulse profiles.

\section{Conclusions and summary}
\label{sec:conclu}
In this work, we have studied the effect of the light bending phenomenon on the signal of a pulsar in a binary system within the full general relativistic framework. We have ignored the spin of the companion, i.e., assumed Schwarzschild geometry of the spacetime near the companion. This choice is justified by studies of stellar evolution. \citet{css21} found that about $80 \%$ of pulsar-black hole systems have the black hole as the first born object, and in that case the spin of the  black hole can be ignored based on the study by \citet{fm19}. Even if the black hole is born second, it would have negligible spin if the value of the orbital period is larger than 2 days (see section 5.2 of \citet{css21}), and in the present work, we have used the value of the orbital period 5.5 days which is the mean value of the orbital period obtained in one of the models of \citet{css21}.

As the bending takes place near the companion, the spin of the pulsar does not play any role in the estimation of the bending. However, it plays a significant role in the geometry, i.e., in determining the initial direction of the pulsar beam. To calculate the initial direction of the beam, in addition to creating the beam at different rotational phases, we also traced the pulsar in its orbit by solving Kepler's equations augmented by post-Keplerian parameters to account for the general relativistic effects. Additionally, we have taken care of the geodetic precession of the spin axis of the pulsar about the total angular momentum of the binary.

The light bending phenomenon results in two different delays, the longitudinal delay and the latitudinal delay. Both delays would be present, but usually one is much stronger than the other depending on the geometry of the system. The resultant bending delay curve, which is the sum of the two delay curves, would mostly follow the shape of the stronger one. Similar to the Shapiro delay (another  manifestation of the spacetime curvature around the companion), the total bending delay is also stronger for edge-on systems, especially at the orbital phases near the superior conjunction. So, it might be difficult to decouple these delays \citep{hkc22}. However, the value of the Shapiro delay depends only on the masses of the pulsar and the companions as well as the orbital parameters, e.g., $e$, $P_b$, $i$, $\omega$, and $A_T$. On the other hand, in addition to these parameters, the values of the angles $\alpha$, $\eta$, and $\lambda$ affect the value of the bending delay. 

Although we have studied the manifestation of the light bending effect within the framework of full general relativity, it is difficult to incorporate this in pulsar timing analysis. However, the simpler analytical expressions given by \citetalias{Rafikov2005} or \citetalias{dk1995} match the full general relativistic methods except some specific situations. The first is the case when the values of both the orbital phase and the inclination angle are close to $90^{\circ}$. In such a case, the bending delay suddenly becomes very strong. The second is the case when the orbital phase is near $270^\circ$, where the expressions of \citetalias{Rafikov2005} gives some artificial features as seen in Fig. \ref{fig:comparison_i_89}. Although the expression by \citetalias{dk1995} is free from any such artificial feature, but they provided the expression for the longitudinal bending delay only, not for the latitudinal bending delay.

In addition to the delay, the phenomenon of bending leads to a distortion of the beam resulting in a change in the pulse profile. This change is significant only when both of the orbital phase and the inclination angle are close to $90^{\circ}$. Timing analysis will be erroneous if one uses this profile as a template to generate ToAs at other orbital phases.  Moreover, the distortion of the beam sometimes results in a tail either at the beginning or at the end of the profiles, as seen in Fig. \ref{pulse_difforbital_phase_i_90}. However, the intensity near such a tail is usually very small (lying in the range of 1 to 5 in our units of intensity) and highly likely to be obscured by the noise. Note that, in Fig. \ref{pulse_phase_90_i_89_7}, we see that the profile variation due to the bending is insignificant for a pulsar$-$black hole binary with $i=89.7^{\circ}$. This also explains the null detection of the bending induced profile variation for the double pulsar for which $i=89.35^{\circ} ~{\rm or}~ 90.65^{\circ}$ \citep{hkc22}. 

It is possible that the first discover pulsar$-$black hole binary will have the orbital inclination around $89^{\circ}$ or less, so that even if the bending delay is present, the pulse shape will not be affected by the bending. Timing such a system will not pose any serious challenges. Even if the system has orbital inclination higher than $89^{\circ}$, the pulse shape will be affected only near the orbital phase of $90^{\circ}$, and timing such a system will be possible by filtering out the data at those orbital phases. Hence, we emphasise the fact that timing a pulsar$-$black hole binary will not be problematic and tests of gravity will be possible.

\section*{Acknowledgements}

The authors thank Maura McLaughlin, A. Gopakumar, and Bhal Chandra Joshi for discussions. They also thank the anonymous referee for constructive suggestions. JD thanks Prateek Chawla for help in coding. AB acknowledges the support from the UK Science and Technology Facilities Council (STFC). A consolidated grant from STFC supports the pulsar research at the Jodrell Bank Centre for Astrophysics.

\section*{Data Availability}

No observational data have been used in this work. Numerical results and codes underlying this article will be shared on reasonable request to the corresponding author.

\bibliographystyle{mnras}
\bibliography{psrbh}

\bsp	
\label{lastpage}
\end{document}